\documentclass[%
 reprint,
superscriptaddress,
nofootinbib,
amsmath,amssymb,
 aps,
]{revtex4-2}

\DeclareUnicodeCharacter{0308}{HERE!HERE!}

\usepackage{lipsum}
\usepackage{xcolor}
\usepackage{mathrsfs,amsmath}
\usepackage{gensymb}
\usepackage{ulem}
\usepackage{xcolor}
\usepackage{graphicx}
\usepackage{dcolumn}
\usepackage{bm}
\usepackage[hidelinks]{hyperref}
\usepackage{miller}
\usepackage{float}
\newcommand{\cmmnt}[1]{\ignorespaces}
\usepackage[compact]{titlesec}
\usepackage[utf8]{inputenc}

\makeatletter
\renewcommand*{\@fnsymbol}[1]{\ensuremath{\ifcase#1\or \dagger \else\@ctrerr\fi}}
\makeatother

\begin{document}

\preprint{APS/123-QED}

\title {A Suspended 4H-Silicon Carbide Membrane Platform \\ for Defect Integration into Quantum Devices}

\author{Amberly H. Xie}
    \thanks{Corresponding Author: axie@g.harvard.edu}
    \affiliation{John A. Paulson School of Engineering and Applied Sciences, Harvard University, Cambridge, Massachusetts 02138, USA}
\author{Aaron M. Day}
    \affiliation{John A. Paulson School of Engineering and Applied Sciences, Harvard University, Cambridge, Massachusetts 02138, USA}
\author{Jonathan R. Dietz}
    \affiliation{IonQ, 1284 Soldiers Fld Rd, Brighton, MA 02135}
\author{Chang Jin}
    \affiliation{John A. Paulson School of Engineering and Applied Sciences, Harvard University, Cambridge, Massachusetts 02138, USA}
\author{Chaoshen Zhang}
    \affiliation{John A. Paulson School of Engineering and Applied Sciences, Harvard University, Cambridge, Massachusetts 02138, USA}
\author{Eliana Mann}
    \affiliation{John A. Paulson School of Engineering and Applied Sciences, Harvard University, Cambridge, Massachusetts 02138, USA}
\author{Zhujing Xu}
    \affiliation{John A. Paulson School of Engineering and Applied Sciences, Harvard University, Cambridge, Massachusetts 02138, USA}
\author{Marko Loncar}
    \affiliation{John A. Paulson School of Engineering and Applied Sciences, Harvard University, Cambridge, Massachusetts 02138, USA}
\author{Evelyn L. Hu}  
    \affiliation{John A. Paulson School of Engineering and Applied Sciences, Harvard University, Cambridge, Massachusetts 02138, USA}

\date{\today}
\begin{abstract}
    4H-silicon carbide is a promising platform for solid-state quantum technology due to its commercial availability as a wide bandgap semiconductor and ability to host numerous spin-active color centers. Integrating color centers into suspended nanodevices enhances defect control and readout--key advances needed to fully harness their potential. However, challenges in developing robust fabrication processes for 4H-SiC thin films--due to the material's chemical and mechanical stability--limit their implementation in quantum applications. Here, we report on a new fabrication approach that first synthesizes suspended thin films from a monolithic platform, then patterns devices. With this technique, we fabricate and characterize structures tailored for defect integration, demonstrating 1D photonic crystal cavities, with and without waveguide interfaces, and lithium niobate on 4H-SiC acoustic cavities. This approach allows for greater fabrication flexibility--supporting high temperature annealing and heterogeneous material platform compatibility--providing a versatile platform for scalable fabrication of 4H-SiC devices for quantum technologies.
\end{abstract}
\maketitle

Solid state point defects can serve as quantum emitters, showing remarkable potential in applications ranging from single-photon emission \cite{castelletto2021silicon, aharonovich2016solid, bathen2021manipulating, khramtsov2021bright}, ultra-precise quantum metrology \cite{dietz2023spin, castelletto2023quantum, giovannetti2001quantum}, and secure quantum networks \cite{knaut2024entanglement, bersin2024telecom, stas2022robust}. To fully leverage their capabilities, many applications require defect integration into devices such as photonic or mechanical resonators \cite{ding2024high, zhou2022photonic} and electronics \cite{sato2018room, day2024electrical} to improve readout \cite{knall2022efficient}, manipulate light emission \cite{wang2021high}, or control defect spin \cite{assumpcao2023deterministic, koller2025strain, meesala2018strain}. A common feature among many of these devices is suspension, which is crucial in ensuring effective mode confinement and strong defect-cavity coupling. As such, the generation of high-quality thin films is a prerequisite which enables current leading quantum technology platforms today, such as silicon and diamond \cite{celler2003frontiers, guo2021tunable, ding2024high}. 

Though many host materials exist for point defect-device integration, 4H-silicon carbide (4H-SiC) in particular stands out due to its exceptional material properties, including large bandgap, large carrier mobility, and high quality wafer-scale epitaxial growth and doping processes \cite{kimoto2014fundamentals}. However, fabrication of 4H-SiC suspended films and devices is challenging due to the material's chemical and mechanical robustness. Various methods have been developed to address this, including integration with on-insulator platforms \cite{lukin20204h}, angled etching \cite{majety2025wafer, song2018high}, and dopant-selective photoelectrochemical (PEC) etching \cite{bracher2015fabrication,bracher2017selective,crook2020purcell}. Despite this, existing implementations of these methods often compromise device performance and yield due to material damage from high-energy ion implantation or extensive wafer grinding \cite{tsukimoto2018local, kozlov2002buried}, restricted device geometries, or reduced emitter spin coherence resulting from a highly doped material stack \cite{bracher2015fabrication, crook2020purcell}. 

\begin{figure*}
    \centering
    \includegraphics[width=1\linewidth]{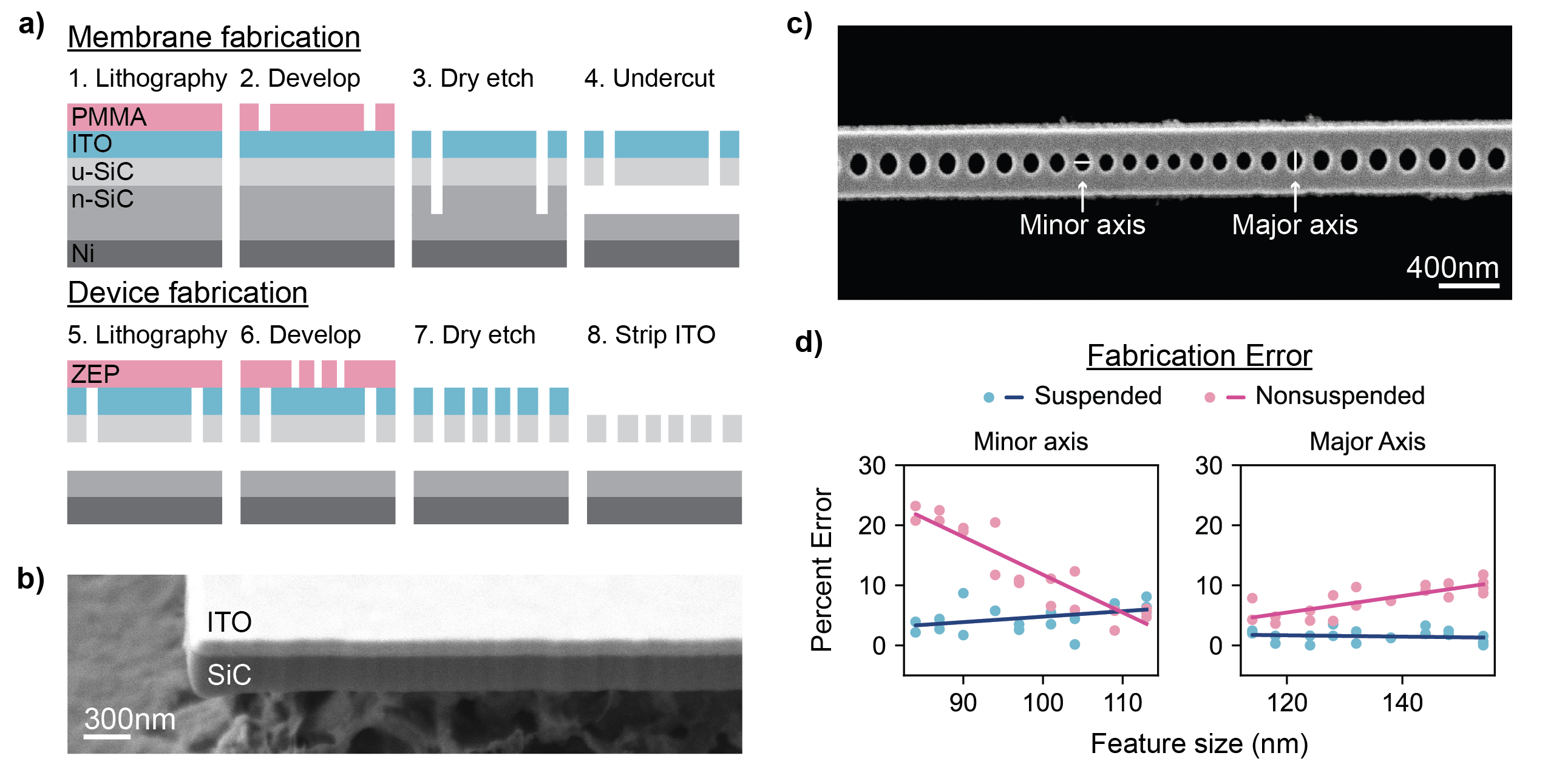}
    \caption{a) Fabrication flow for suspended membrane generation and subsequent device patterning and etching. u-SiC (unintentionally doped-SiC) represents the low-doped epitaxial layer. b) SEM sideview of suspended membrane. Corresponds to step 4 in (a). c) SEM of 1D PhC defining major and minor axes. d) Comparison of percent error for features features onto suspended vs. nonsuspended substrates. Percent error defined as: (target feature size - fabricated feature size)/target feature size. Data taken by measuring holes from 1D PhCs written without any proximity effect correction. Doses chosen independently for suspended and nonsuspended devices to minimize percent error.}
    \label{Fig1}
\end{figure*}

In previous work, we addressed these challenges by introducing a novel PEC etching method for forming suspended 4H-SiC thin films ($<$500 nm) in an undoped epitaxial layer \cite{dietz2025selective}. Here, we present a new technique for fabricating photonic and phononic devices where large membranes are first suspended using this new PEC protocol, onto which devices are subsequently patterned and etched. In contrast to traditional fabrication approaches, where devices are first defined in unsuspended material and then undercut, our approach offers several advantages. We show through the fabrication of 1D photonic crystal cavities (PhCs) that patterning on suspended membranes decreases secondary electron production and proximity effect induced fabrication errors. Furthermore, we demonstrate the extensibility of our platform by fabricating tapered fiber coupled cavities and a suspended thin film lithium niobate (TFLN)-on-SiC acoustic resonator, advancing the platform for point defect integration into suspended 4H-SiC based devices.

The sample stack consisted of 500 nm of unintentionally-doped (1e14 $cm^{-3}$ p-type, Xiamen Powerway, 1e15 $cm^{-3}$ n-type, Wolfspeed) grown epitaxially on n-type (1e18 $cm^{-3}$ Xiamen Powerway, 7e18 $cm^{-3}$ Wolfspeed) $4^\circ$ cut c-axis 4H-SiC wafers. The following fabrication steps were based from \cite{dietz2025integrating}. Samples were cleaned with piranha solution (3:1 $H_2SO_4$/30\% $H_2O_2$) and 49\% HF. A 150 nm layer of nickel, used as a backside contact for PEC, was electron beam evaporated onto the n-type side and annealed using rapid thermal annealing ($900^\circ$C, 3 min, Ar). Samples were re-cleaned and 100 nm of indium tin oxide (ITO), used as a hard mask and top contact for PEC, was sputtered on (base pressure $<$ 2e-6 torr, 600$^\circ$C) the epilayer side. Membrane etch trenches were patterned using PMMA C6 (Kayaku Advanced Materials Inc, 3k rpm, 90$^\circ$C for 3 min then 180$^\circ$C for 7 min, 1:3 MIBK:IPA for 70 seconds) and electron beam lithography (Elionix F125, 125kV, 2nA, 1400 $uC/cm^2$). Reactive ion etching transferred the pattern first to the ITO hard mask (10 mtorr, 30 sccm $H_2$:10 sccm $CF_4$, 400W ICP, 100W RF, 40$^\circ$C) then into the SiC substrate (5 mtorr, 40 sccm $SF_6$:10 sccm $O_2$, 400W ICP, 100W RF, 40$^\circ$C). 

Suspended 4H-SiC thin films were fabricated using a modified version of PEC etching (Figure \ref{Fig1}a) \cite{dietz2025selective}. This approach enables direct suspension of undoped thin films from monolithic wafers without the need for wafer bonding, grinding, or chemical mechanical polishing. 

Under alkaline conditions and above bandgap illumination, SiC can undergo dissolution. This proceeds via a two-step reaction of oxide formation (Equation 1) followed by dissolution (Equation 2). 
\begin{equation}
    SiC + 6OH^- +6h^+ \rightarrow SiO_2 + CO + 3H_2O
\end{equation}
\begin{equation}
    SiO_2 + 2OH^- \rightarrow [Si(OH)_2O_2]^{2-}
\end{equation}

Due to dopant-dependent band bending at the semiconductor-electrolyte interface \cite{zhang2012band, dietz2025selective, pavunny2019doping}, holes migrate towards the electrolyte in n-doped material and contribute to etching, but stay confined within and thus preserve p-doped material. Additionally, we introduce two external biases to improve the undercutting efficiency and final quality of the fabricated thin films:

\begin{enumerate}
    \item A substrate/electrolyte bias (via a Pt counter electrode) to promote hole migration from the n-type substrate to the electrolyte 
    \item An epilayer/substrate reverse bias for etch selectivity against the unintentionally-doped epilayer
\end{enumerate}

Hydroxide ions were provided by an aqueous solution of KOH (50-100 mM) which was flowed continuously through the PEC cell via a peristaltic pump. Holes were generated by above bandgap UV illumination (500W Newport Hg/Xe lamp). Below bandgap light was filtered out using a colored glass filter to prevent electrolyte and sample heating. Sample electrodes were contacted using copper tape. 

Suspended membrane samples (Figure \ref{Fig1}b) were then solvent cleaned (acetone/IPA) and patterned with ZEP520a (Zeon Speciality Materials Inc., 4k rpm, 115$^\circ$C for 3.5 min, o-xylene for 1 min) and espacer (2k rpm, water for 1 min). Electron beam lithography was carried out using Elionix F150 (150kV, 200-300pA, 400-600 $uC/cm^2$). The same dry etch chemistry as stated previously was utilized. ITO was removed using 49\% HF (2x 2 min, for oxide removal) followed by HCl (1 min, for residual In and Sn removal). We found that lowering the baking temperature and spinning the samples on a carrier wafer provided the most reliable method for successful resist application to avoid delamination and adhesion issues (see Supplemental Information for further details). 

One major benefit to fabricating devices directly onto suspended films is mitigation of the proximity effect during lithography due to reduced electron backscattering. To quantify this advantage, we measured and compared the percent error for features fabricated without any proximity effect correction on suspended and nonsuspended films (Figure \ref{Fig1}d). Fabrication onto suspended films not only yielded lower percent errors, but also percent errors with reduced sensitivity to variations in feature size and orientation. Because of this, the lithography optimization process to achieve desired feature sizes is simplified compared with patterning on nonsuspended substrates. The observed discrepancy in percent error between minor and major axes on nonsuspended substrates is attributed to the electron beam rastering horizontally during exposure (parallel to minor axis), causing relative overexposure of horizontal features. 

To showcase the potential of this fabrication approach, we began by fabricating and characterizing baseline 1D photonic crystal cavities (PhC) with and without tapered waveguide interfaces. PhCs are a fundamental component of integrating defects into photonic systems by enhancing zero phonon line (ZPL) emission through the Purcell effect, an essential requirement for many of the aforementioned quantum applications \cite{bracher2017selective}. For instance, ZPL emission provides indistinguishable photons which are needed for various quantum computing schemes \cite{aharonovich2016solid} and high-precision metrology \cite{giovannetti2011advances}.

\begin{figure}[ht!]
    \centering
    \includegraphics[width=1\linewidth]{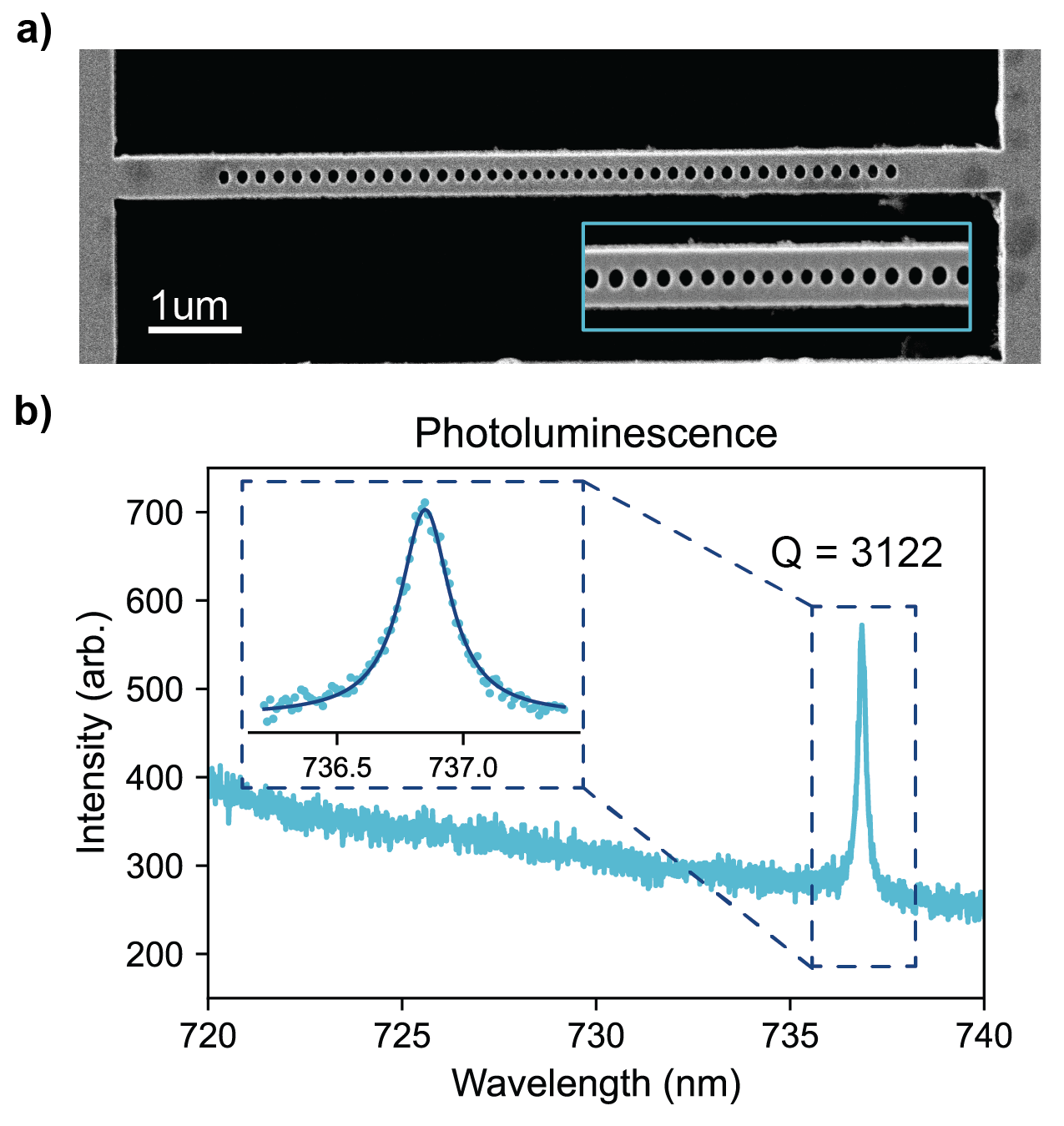}
    \caption{a) SEM of 1D PhC fabricated by direct patterning onto suspended membranes. b) Room temperature photoluminescence data measuring cavity mode. Cavity mode fit to a Lorentzian peak.}
    \label{Fig2}
\end{figure}

1D PhCs (Figure \ref{Fig2}) were designed to be resonant with surface states (650 - 750 nm), intrinsic defects that arise from a combination of dangling surface bonds and surface oxide, for ease of device characterization (see \cite{SiCNBcode} for simulation details) \cite{bracher2015fabrication, sato2018room, day2023laser}. We characterized the cavities through room temperature photoluminescence (see Supplemental Information) and measured quality factors of a few thousand (Figure \ref{Fig2}b). These first cavity Q results are comparable with other 4H-SiC PhCs reported in literature \cite{bracher2015fabrication, zhou2022photonic}, confirming the efficacy of direct writing and etching onto suspended membranes. Because these first studies matched the resonance of the PhC to prevalent surface states, we believe that their absorption limits the observed Q and further optimization of this process combined with defect integration will result in devices with high defect-cavity coupling.

\begin{figure}[hbt!]
    \centering
    \includegraphics[width=1\linewidth]{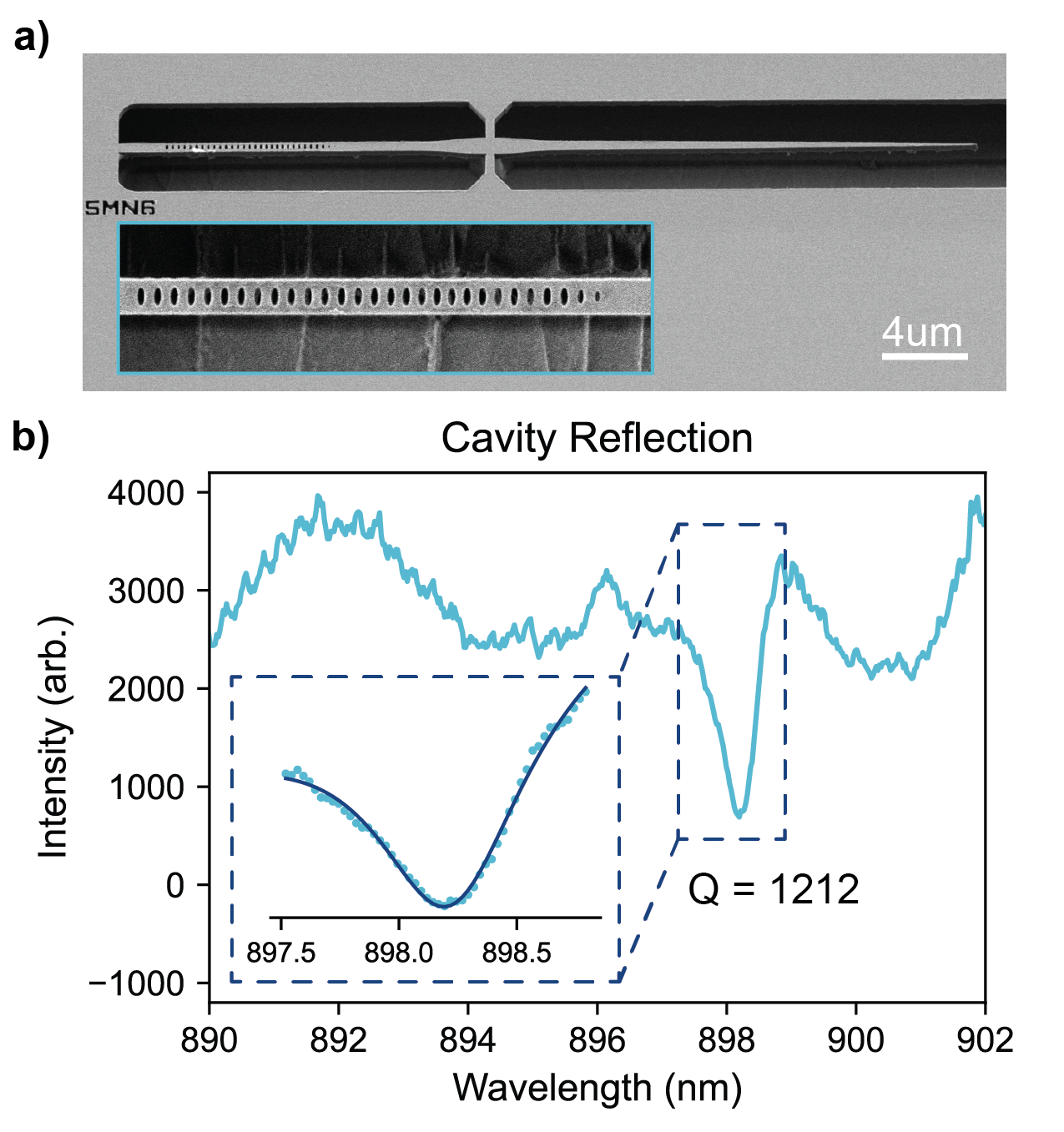}
    \caption{a) SEM of tapered waveguide interfaced cavity fabricated onto suspended membranes. b) Cavity reflection measurements showing resonant dip at cavity mode. Inset shows reflection dip fitted to a Lorentzian with a linear background to account for fringes from excitation light.}
    \label{Fig3}
\end{figure}

\begin{figure*}[hbt!]
    \centering
    \includegraphics[width=1\linewidth]{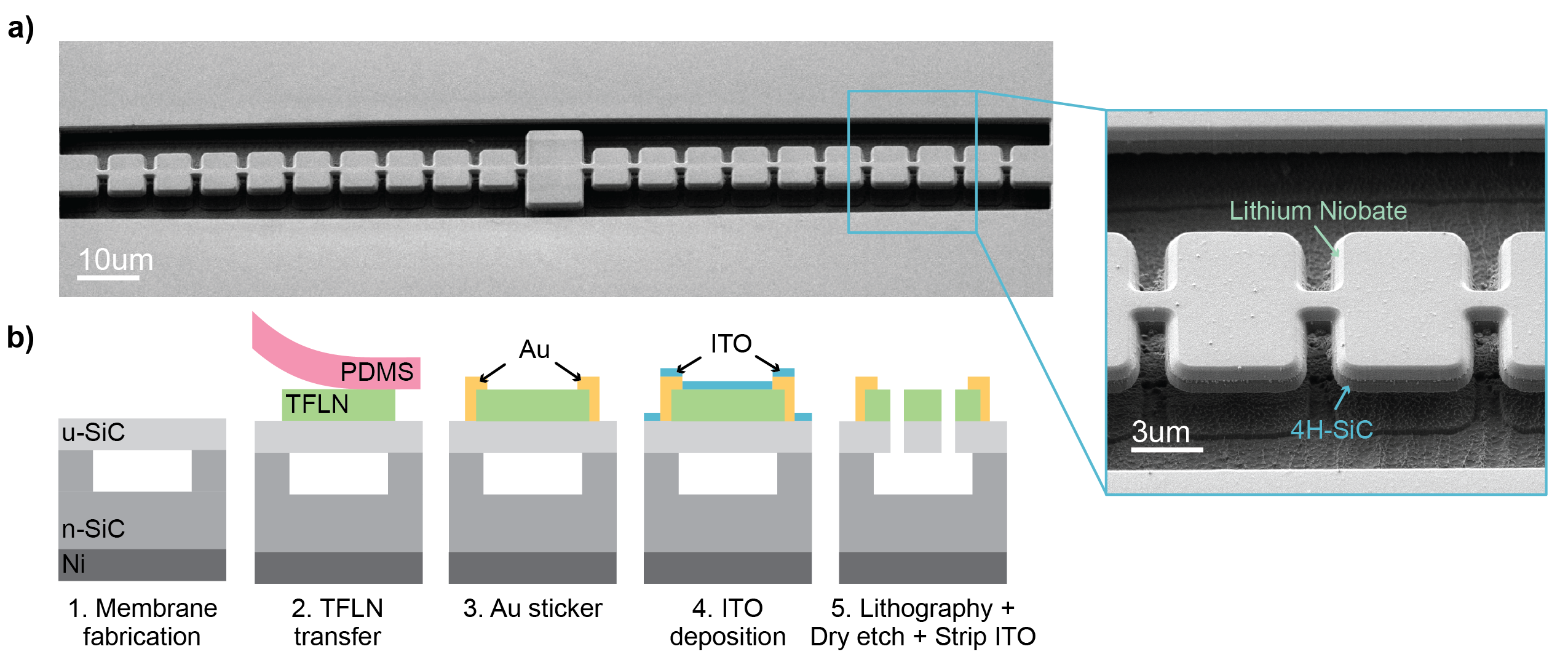}
    \caption{a) SEM of TFLN-SiC block tether cavity. Right image shows individual layers of TFLN and SiC. b) Fabrication flow of block tether cavities.}
    \label{Fig4}
\end{figure*}

One major limitation of the aforementioned PhCs is inefficient light collection due to reliance on free-space scattering for photon extraction. However, this can be mitigated by integrating a tapered waveguide region to the end of the PhCs, directing light collection and improving readout efficiency through coupling to optical fibers \cite{knall2022efficient, ding2024high}. Furthermore, fiber coupling supports capabilities such as defect-defect interaction and multinode entanglement by providing a pathway for photon transport between PhCs \cite{knaut2024entanglement}. Thus, we fabricated tapered fiber cavities with resonant modes targeted around $V_{Si}^-$ emission (see \cite{SiCTCcode} for simulation details) utilizing our newly developed approach to support fiber coupling and enhance readout efficiency (Figure \ref{Fig3}a). 

We targeted critically coupled cavities to facilitate ease of measurement. Reflection measurements were performed to assess cavity performance, yielding quality factors Q$>$1,000. Though further optimization would be beneficial, the current results are sufficient to support continued development of this fabrication technique.  Integrating nanophotonic resonators with optical fibers is essential in realizing a scalable quantum network, enabling long-distance quantum information distribution \cite{bersin2024development, bersin2024telecom}. Thus, these demonstrated devices further solidify silicon carbide as a competitive platform for advancing quantum networking applications.

Integration into mechanical devices offers additional degrees of freedom for point defect coupling in which optical control and readout are insufficient. Such spin-phonon interactions enable access to additional ground state spin-level transitions that are optically forbidden \cite{dietz2023spin, vasselon2023acoustically}, allow for interfacing between different qubit platforms \cite{schutz2017universal}, and enable electrical-based readout and control of defects with piezoelectric tuning through phonon-mediated transduction \cite{chen2020acoustically}. These capabilities hold significant potential for color center control and readout in 4H-SiC. For example, the metastable state present within the $V_{Si}^-$ level structure reduces the efficiency of optical based readout \cite{soykal2016silicon} and only certain ground state spin transitions are allowed optically \cite{dietz2023spin}. However, this limitation could be resolved through mechanical coupling with some piezoelectric platform, facilitating efficient readout through piezoelectric transduction and access to more ground state spin level transitions. 

To this end, we fabricated a phononic crystal cavity integrating thin film lithium niobate (TFLN) with our new fabrication process (Figure \ref{Fig4}a), demonstrating the versatility of this platform with heterogenous material integration to obtain fully suspended TFLN-on-SiC. Lithium niobate was chosen because of its large electromechanical coupling factor \cite{olsson2014high}, allowing for strong acoustic wave transduction into the SiC and electric-based readout mediated by spin-phonon coupling. 

For device fabrication (Figure \ref{Fig4}b), u-SiC membranes were first generated using the previously described methods. A 600 nm thick x-cut TFLN layer was transferred onto the SiC utilizing a PDMS transfer process \cite{xu2025thin}. Electron beam lithography was used to define a frame around the TFLN border for deposition and liftoff of a Au sticker to secure the TFLN on the carrier chip \cite{ding2024high}. Fresh ITO was deposited for use as a hard mask and the same procedure as outlined previously was used for device patterning. For pattern transfer, the sample stack was etched in three steps: ITO etching, argon ion etching for pattern transfer into the TFLN, and SiC etching. Lastly, ITO was removed using HF and HCl. See Supplemental Information for simulation details. 

Direct fabrication on suspended SiC membranes eliminated the need for developing post-processing steps for undercutting the SiC layer, allowing us to preserve the integrity of the remaining material stack. While here we demonstrated a proof-of-concept device for heterogeneous integration of TFLN-on-SiC, additional development is required to enable practical spin control and readout of color centers. Additionally, this platform offers promising future opportunities for more complex point defect applications such as electro-optic modulation of emitted photons \cite{li2020lithium}, in-situ cavity tuning into resonance with the defect ZPL \cite{xia2022}, and multiplexing quantum nodes \cite{assumpcao2024thin}.

In summary, we have developed and report on a new fabrication approach for 4H-SiC devices where we generate suspended thin-films of SiC and then proceed with device patterning. By patterning directly onto suspended membranes, it was found that the decrease in proximity effect resulted in lower percent error for fabricated feature dimensions and improved consistency across feature size and orientation. This method of direct patterning onto suspended films can be extended to other thin-film material platforms as well for quantum emitter integration into devices.

We showcased the viability of this platform by fabricating and characterizing initial baseline photonic devices--free-space scattering and tapered waveguide-coupled 1D PhCs--achieving quality factors of a few thousand. While tapered waveguide interfaces have been recently demonstrated \cite{krumrein2024precise}, to the best of our knowledge, these represent the first reported demonstrations of tapered waveguide cavities in 4H-SiCs. To illustrate the potential of this approach for realizing complex devices, we additionally fabricated an acoustic resonator integrating TFLN with our suspended SiC platform, enabling capabilities such as enhanced spin-state readout through piezoelectric transduction and surface acoustic wave generation for color center control \cite{campbell202421}. 

Additionally, this platform is more robust to high temperature fabrication steps compared with other thin film SiC platforms. Unlike composite material stacks, which suffer from thermal expansion coefficient mismatch, the suspended films reported on here can withstand high temperature annealing due to their monolithic nature \cite{hochreiter2023electrochemical}. This enables a broader range of devices for defect integration, accommodating electrical devices that require high temperature annealing for dopant activation \cite{das2024study, negoro2004electrical}, and exploration of telecom emitting transition metal defects, such as vanadium, which require high-temperature annealing following defect implantation \cite{wolfowicz2020vanadium}. The results reported on here demonstrate that this newly developed fabrication approach offers a flexible and robust 4H-SiC material platform for future complex device synthesis and defect integration. 

\section{Supporting Information}
Additional methods regarding direct fabrication onto membranes, computational design of devices, and photoluminescence measurements methods.

\section{Acknowledgments}
\noindent The authors thank Chawina De-Eknamkul for her help in fabricating tapered fibers for characterizing the tapered cavities and Flexcompute Inc. for providing the Tidy3D software used to design the photonic crystal cavities described here. This work was supported by AWS Center for Quantum Networking and the Harvard Quantum Initiative. Portions of this work were performed at the Harvard University Center for Nanoscale Systems (CNS); a member of the National Nanotechnology Coordinated Infrastructure Network (NNCI), which is supported by the National Science Foundation under NSF award no. ECCS-2025158.

\section{Data Availability}
\noindent The data that support the findings of the work are available from the corresponding author upon reasonable request. 

\normalem
\bibliography{Ref.bib}

\begin{thebibliography}{54}%
\makeatletter
\providecommand \@ifxundefined [1]{%
 \@ifx{#1\undefined}
}%
\providecommand \@ifnum [1]{%
 \ifnum #1\expandafter \@firstoftwo
 \else \expandafter \@secondoftwo
 \fi
}%
\providecommand \@ifx [1]{%
 \ifx #1\expandafter \@firstoftwo
 \else \expandafter \@secondoftwo
 \fi
}%
\providecommand \natexlab [1]{#1}%
\providecommand \enquote  [1]{``#1''}%
\providecommand \bibnamefont  [1]{#1}%
\providecommand \bibfnamefont [1]{#1}%
\providecommand \citenamefont [1]{#1}%
\providecommand \href@noop [0]{\@secondoftwo}%
\providecommand \href [0]{\begingroup \@sanitize@url \@href}%
\providecommand \@href[1]{\@@startlink{#1}\@@href}%
\providecommand \@@href[1]{\endgroup#1\@@endlink}%
\providecommand \@sanitize@url [0]{\catcode `\\12\catcode `\$12\catcode `\&12\catcode `\#12\catcode `\^12\catcode `\_12\catcode `\%12\relax}%
\providecommand \@@startlink[1]{}%
\providecommand \@@endlink[0]{}%
\providecommand \url  [0]{\begingroup\@sanitize@url \@url }%
\providecommand \@url [1]{\endgroup\@href {#1}{\urlprefix }}%
\providecommand \urlprefix  [0]{URL }%
\providecommand \Eprint [0]{\href }%
\providecommand \doibase [0]{https://doi.org/}%
\providecommand \selectlanguage [0]{\@gobble}%
\providecommand \bibinfo  [0]{\@secondoftwo}%
\providecommand \bibfield  [0]{\@secondoftwo}%
\providecommand \translation [1]{[#1]}%
\providecommand \BibitemOpen [0]{}%
\providecommand \bibitemStop [0]{}%
\providecommand \bibitemNoStop [0]{.\EOS\space}%
\providecommand \EOS [0]{\spacefactor3000\relax}%
\providecommand \BibitemShut  [1]{\csname bibitem#1\endcsname}%
\let\auto@bib@innerbib\@empty
\bibitem [{\citenamefont {Castelletto}(2021)}]{castelletto2021silicon}%
  \BibitemOpen
  \bibfield  {author} {\bibinfo {author} {\bibfnamefont {S.}~\bibnamefont {Castelletto}},\ }\bibfield  {title} {\bibinfo {title} {{Silicon carbide single-photon sources: challenges and prospects}},\ }\href@noop {} {\bibfield  {journal} {\bibinfo  {journal} {Materials for Quantum Technology}\ }\textbf {\bibinfo {volume} {1}},\ \bibinfo {pages} {023001} (\bibinfo {year} {2021})}\BibitemShut {NoStop}%
\bibitem [{\citenamefont {Aharonovich}\ \emph {et~al.}(2016)\citenamefont {Aharonovich}, \citenamefont {Englund},\ and\ \citenamefont {Toth}}]{aharonovich2016solid}%
  \BibitemOpen
  \bibfield  {author} {\bibinfo {author} {\bibfnamefont {I.}~\bibnamefont {Aharonovich}}, \bibinfo {author} {\bibfnamefont {D.}~\bibnamefont {Englund}},\ and\ \bibinfo {author} {\bibfnamefont {M.}~\bibnamefont {Toth}},\ }\bibfield  {title} {\bibinfo {title} {Solid-state single-photon emitters},\ }\href@noop {} {\bibfield  {journal} {\bibinfo  {journal} {Nature photonics}\ }\textbf {\bibinfo {volume} {10}},\ \bibinfo {pages} {631} (\bibinfo {year} {2016})}\BibitemShut {NoStop}%
\bibitem [{\citenamefont {Bathen}\ and\ \citenamefont {Vines}(2021)}]{bathen2021manipulating}%
  \BibitemOpen
  \bibfield  {author} {\bibinfo {author} {\bibfnamefont {M.~E.}\ \bibnamefont {Bathen}}\ and\ \bibinfo {author} {\bibfnamefont {L.}~\bibnamefont {Vines}},\ }\bibfield  {title} {\bibinfo {title} {Manipulating single-photon emission from point defects in diamond and silicon carbide},\ }\href@noop {} {\bibfield  {journal} {\bibinfo  {journal} {Advanced Quantum Technologies}\ }\textbf {\bibinfo {volume} {4}},\ \bibinfo {pages} {2100003} (\bibinfo {year} {2021})}\BibitemShut {NoStop}%
\bibitem [{\citenamefont {Khramtsov}\ and\ \citenamefont {Fedyanin}(2021)}]{khramtsov2021bright}%
  \BibitemOpen
  \bibfield  {author} {\bibinfo {author} {\bibfnamefont {I.~A.}\ \bibnamefont {Khramtsov}}\ and\ \bibinfo {author} {\bibfnamefont {D.~Y.}\ \bibnamefont {Fedyanin}},\ }\bibfield  {title} {\bibinfo {title} {Bright silicon carbide single-photon emitting diodes at low temperatures: Toward quantum photonics applications},\ }\href@noop {} {\bibfield  {journal} {\bibinfo  {journal} {Nanomaterials}\ }\textbf {\bibinfo {volume} {11}},\ \bibinfo {pages} {3177} (\bibinfo {year} {2021})}\BibitemShut {NoStop}%
\bibitem [{\citenamefont {Dietz}\ \emph {et~al.}(2023)\citenamefont {Dietz}, \citenamefont {Jiang}, \citenamefont {Day}, \citenamefont {Bhave},\ and\ \citenamefont {Hu}}]{dietz2023spin}%
  \BibitemOpen
  \bibfield  {author} {\bibinfo {author} {\bibfnamefont {J.~R.}\ \bibnamefont {Dietz}}, \bibinfo {author} {\bibfnamefont {B.}~\bibnamefont {Jiang}}, \bibinfo {author} {\bibfnamefont {A.~M.}\ \bibnamefont {Day}}, \bibinfo {author} {\bibfnamefont {S.~A.}\ \bibnamefont {Bhave}},\ and\ \bibinfo {author} {\bibfnamefont {E.~L.}\ \bibnamefont {Hu}},\ }\bibfield  {title} {\bibinfo {title} {Spin-acoustic control of silicon vacancies in 4h silicon carbide},\ }\href@noop {} {\bibfield  {journal} {\bibinfo  {journal} {Nature Electronics}\ }\textbf {\bibinfo {volume} {6}},\ \bibinfo {pages} {739} (\bibinfo {year} {2023})}\BibitemShut {NoStop}%
\bibitem [{\citenamefont {Castelletto}\ \emph {et~al.}(2023)\citenamefont {Castelletto}, \citenamefont {Lew}, \citenamefont {Lin},\ and\ \citenamefont {Xu}}]{castelletto2023quantum}%
  \BibitemOpen
  \bibfield  {author} {\bibinfo {author} {\bibfnamefont {S.}~\bibnamefont {Castelletto}}, \bibinfo {author} {\bibfnamefont {C.~T.}\ \bibnamefont {Lew}}, \bibinfo {author} {\bibfnamefont {W.-X.}\ \bibnamefont {Lin}},\ and\ \bibinfo {author} {\bibfnamefont {J.-S.}\ \bibnamefont {Xu}},\ }\bibfield  {title} {\bibinfo {title} {Quantum systems in silicon carbide for sensing applications},\ }\href@noop {} {\bibfield  {journal} {\bibinfo  {journal} {Reports on Progress in Physics}\ }\textbf {\bibinfo {volume} {87}},\ \bibinfo {pages} {014501} (\bibinfo {year} {2023})}\BibitemShut {NoStop}%
\bibitem [{\citenamefont {Giovannetti}\ \emph {et~al.}(2001)\citenamefont {Giovannetti}, \citenamefont {Lloyd},\ and\ \citenamefont {Maccone}}]{giovannetti2001quantum}%
  \BibitemOpen
  \bibfield  {author} {\bibinfo {author} {\bibfnamefont {V.}~\bibnamefont {Giovannetti}}, \bibinfo {author} {\bibfnamefont {S.}~\bibnamefont {Lloyd}},\ and\ \bibinfo {author} {\bibfnamefont {L.}~\bibnamefont {Maccone}},\ }\bibfield  {title} {\bibinfo {title} {Quantum-enhanced positioning and clock synchronization},\ }\href@noop {} {\bibfield  {journal} {\bibinfo  {journal} {Nature}\ }\textbf {\bibinfo {volume} {412}},\ \bibinfo {pages} {417} (\bibinfo {year} {2001})}\BibitemShut {NoStop}%
\bibitem [{\citenamefont {Knaut}\ \emph {et~al.}(2024)\citenamefont {Knaut}, \citenamefont {Suleymanzade}, \citenamefont {Wei}, \citenamefont {Assumpcao}, \citenamefont {Stas}, \citenamefont {Huan}, \citenamefont {Machielse}, \citenamefont {Knall}, \citenamefont {Sutula}, \citenamefont {Baranes} \emph {et~al.}}]{knaut2024entanglement}%
  \BibitemOpen
  \bibfield  {author} {\bibinfo {author} {\bibfnamefont {C.~M.}\ \bibnamefont {Knaut}}, \bibinfo {author} {\bibfnamefont {A.}~\bibnamefont {Suleymanzade}}, \bibinfo {author} {\bibfnamefont {Y.-C.}\ \bibnamefont {Wei}}, \bibinfo {author} {\bibfnamefont {D.~R.}\ \bibnamefont {Assumpcao}}, \bibinfo {author} {\bibfnamefont {P.-J.}\ \bibnamefont {Stas}}, \bibinfo {author} {\bibfnamefont {Y.~Q.}\ \bibnamefont {Huan}}, \bibinfo {author} {\bibfnamefont {B.}~\bibnamefont {Machielse}}, \bibinfo {author} {\bibfnamefont {E.~N.}\ \bibnamefont {Knall}}, \bibinfo {author} {\bibfnamefont {M.}~\bibnamefont {Sutula}}, \bibinfo {author} {\bibfnamefont {G.}~\bibnamefont {Baranes}}, \emph {et~al.},\ }\bibfield  {title} {\bibinfo {title} {Entanglement of nanophotonic quantum memory nodes in a telecom network},\ }\href@noop {} {\bibfield  {journal} {\bibinfo  {journal} {Nature}\ }\textbf {\bibinfo {volume} {629}},\ \bibinfo {pages} {573} (\bibinfo {year} {2024})}\BibitemShut {NoStop}%
\bibitem [{\citenamefont {Bersin}\ \emph {et~al.}(2024{\natexlab{a}})\citenamefont {Bersin}, \citenamefont {Sutula}, \citenamefont {Huan}, \citenamefont {Suleymanzade}, \citenamefont {Assumpcao}, \citenamefont {Wei}, \citenamefont {Stas}, \citenamefont {Knaut}, \citenamefont {Knall}, \citenamefont {Langrock} \emph {et~al.}}]{bersin2024telecom}%
  \BibitemOpen
  \bibfield  {author} {\bibinfo {author} {\bibfnamefont {E.}~\bibnamefont {Bersin}}, \bibinfo {author} {\bibfnamefont {M.}~\bibnamefont {Sutula}}, \bibinfo {author} {\bibfnamefont {Y.~Q.}\ \bibnamefont {Huan}}, \bibinfo {author} {\bibfnamefont {A.}~\bibnamefont {Suleymanzade}}, \bibinfo {author} {\bibfnamefont {D.~R.}\ \bibnamefont {Assumpcao}}, \bibinfo {author} {\bibfnamefont {Y.-C.}\ \bibnamefont {Wei}}, \bibinfo {author} {\bibfnamefont {P.-J.}\ \bibnamefont {Stas}}, \bibinfo {author} {\bibfnamefont {C.~M.}\ \bibnamefont {Knaut}}, \bibinfo {author} {\bibfnamefont {E.~N.}\ \bibnamefont {Knall}}, \bibinfo {author} {\bibfnamefont {C.}~\bibnamefont {Langrock}}, \emph {et~al.},\ }\bibfield  {title} {\bibinfo {title} {Telecom networking with a diamond quantum memory},\ }\href@noop {} {\bibfield  {journal} {\bibinfo  {journal} {PRX Quantum}\ }\textbf {\bibinfo {volume} {5}},\ \bibinfo {pages} {010303} (\bibinfo {year} {2024}{\natexlab{a}})}\BibitemShut {NoStop}%
\bibitem [{\citenamefont {Stas}\ \emph {et~al.}(2022)\citenamefont {Stas}, \citenamefont {Huan}, \citenamefont {Machielse}, \citenamefont {Knall}, \citenamefont {Suleymanzade}, \citenamefont {Pingault}, \citenamefont {Sutula}, \citenamefont {Ding}, \citenamefont {Knaut}, \citenamefont {Assumpcao} \emph {et~al.}}]{stas2022robust}%
  \BibitemOpen
  \bibfield  {author} {\bibinfo {author} {\bibfnamefont {P.-J.}\ \bibnamefont {Stas}}, \bibinfo {author} {\bibfnamefont {Y.~Q.}\ \bibnamefont {Huan}}, \bibinfo {author} {\bibfnamefont {B.}~\bibnamefont {Machielse}}, \bibinfo {author} {\bibfnamefont {E.~N.}\ \bibnamefont {Knall}}, \bibinfo {author} {\bibfnamefont {A.}~\bibnamefont {Suleymanzade}}, \bibinfo {author} {\bibfnamefont {B.}~\bibnamefont {Pingault}}, \bibinfo {author} {\bibfnamefont {M.}~\bibnamefont {Sutula}}, \bibinfo {author} {\bibfnamefont {S.~W.}\ \bibnamefont {Ding}}, \bibinfo {author} {\bibfnamefont {C.~M.}\ \bibnamefont {Knaut}}, \bibinfo {author} {\bibfnamefont {D.~R.}\ \bibnamefont {Assumpcao}}, \emph {et~al.},\ }\bibfield  {title} {\bibinfo {title} {Robust multi-qubit quantum network node with integrated error detection},\ }\href@noop {} {\bibfield  {journal} {\bibinfo  {journal} {Science}\ }\textbf {\bibinfo {volume} {378}},\ \bibinfo {pages} {557} (\bibinfo {year} {2022})}\BibitemShut {NoStop}%
\bibitem [{\citenamefont {Ding}\ \emph {et~al.}(2024)\citenamefont {Ding}, \citenamefont {Haas}, \citenamefont {Guo}, \citenamefont {Kuruma}, \citenamefont {Jin}, \citenamefont {Li}, \citenamefont {Awschalom}, \citenamefont {Delegan}, \citenamefont {Heremans}, \citenamefont {High} \emph {et~al.}}]{ding2024high}%
  \BibitemOpen
  \bibfield  {author} {\bibinfo {author} {\bibfnamefont {S.~W.}\ \bibnamefont {Ding}}, \bibinfo {author} {\bibfnamefont {M.}~\bibnamefont {Haas}}, \bibinfo {author} {\bibfnamefont {X.}~\bibnamefont {Guo}}, \bibinfo {author} {\bibfnamefont {K.}~\bibnamefont {Kuruma}}, \bibinfo {author} {\bibfnamefont {C.}~\bibnamefont {Jin}}, \bibinfo {author} {\bibfnamefont {Z.}~\bibnamefont {Li}}, \bibinfo {author} {\bibfnamefont {D.~D.}\ \bibnamefont {Awschalom}}, \bibinfo {author} {\bibfnamefont {N.}~\bibnamefont {Delegan}}, \bibinfo {author} {\bibfnamefont {F.~J.}\ \bibnamefont {Heremans}}, \bibinfo {author} {\bibfnamefont {A.~A.}\ \bibnamefont {High}}, \emph {et~al.},\ }\bibfield  {title} {\bibinfo {title} {High-q cavity interface for color centers in thin film diamond},\ }\href@noop {} {\bibfield  {journal} {\bibinfo  {journal} {nature communications}\ }\textbf {\bibinfo {volume} {15}},\ \bibinfo {pages} {6358} (\bibinfo {year} {2024})}\BibitemShut {NoStop}%
\bibitem [{\citenamefont {Zhou}\ \emph {et~al.}(2022)\citenamefont {Zhou}, \citenamefont {Wang}, \citenamefont {Yi}, \citenamefont {Shen}, \citenamefont {Zhu}, \citenamefont {Huang}, \citenamefont {Zhou}, \citenamefont {Zhang},\ and\ \citenamefont {Ou}}]{zhou2022photonic}%
  \BibitemOpen
  \bibfield  {author} {\bibinfo {author} {\bibfnamefont {L.}~\bibnamefont {Zhou}}, \bibinfo {author} {\bibfnamefont {C.}~\bibnamefont {Wang}}, \bibinfo {author} {\bibfnamefont {A.}~\bibnamefont {Yi}}, \bibinfo {author} {\bibfnamefont {C.}~\bibnamefont {Shen}}, \bibinfo {author} {\bibfnamefont {Y.}~\bibnamefont {Zhu}}, \bibinfo {author} {\bibfnamefont {K.}~\bibnamefont {Huang}}, \bibinfo {author} {\bibfnamefont {M.}~\bibnamefont {Zhou}}, \bibinfo {author} {\bibfnamefont {J.}~\bibnamefont {Zhang}},\ and\ \bibinfo {author} {\bibfnamefont {X.}~\bibnamefont {Ou}},\ }\bibfield  {title} {\bibinfo {title} {Photonic crystal nanobeam cavities based on 4h-silicon carbide on insulator},\ }\href@noop {} {\bibfield  {journal} {\bibinfo  {journal} {Chinese Optics Letters}\ }\textbf {\bibinfo {volume} {20}},\ \bibinfo {pages} {031302} (\bibinfo {year} {2022})}\BibitemShut {NoStop}%
\bibitem [{\citenamefont {Sato}\ \emph {et~al.}(2018)\citenamefont {Sato}, \citenamefont {Honda}, \citenamefont {Makino}, \citenamefont {Hijikata}, \citenamefont {Lee},\ and\ \citenamefont {Ohshima}}]{sato2018room}%
  \BibitemOpen
  \bibfield  {author} {\bibinfo {author} {\bibfnamefont {S.-i.}\ \bibnamefont {Sato}}, \bibinfo {author} {\bibfnamefont {T.}~\bibnamefont {Honda}}, \bibinfo {author} {\bibfnamefont {T.}~\bibnamefont {Makino}}, \bibinfo {author} {\bibfnamefont {Y.}~\bibnamefont {Hijikata}}, \bibinfo {author} {\bibfnamefont {S.-Y.}\ \bibnamefont {Lee}},\ and\ \bibinfo {author} {\bibfnamefont {T.}~\bibnamefont {Ohshima}},\ }\bibfield  {title} {\bibinfo {title} {Room temperature electrical control of single photon sources at 4h-sic surface},\ }\href@noop {} {\bibfield  {journal} {\bibinfo  {journal} {ACS Photonics}\ }\textbf {\bibinfo {volume} {5}},\ \bibinfo {pages} {3159} (\bibinfo {year} {2018})}\BibitemShut {NoStop}%
\bibitem [{\citenamefont {Day}\ \emph {et~al.}(2024)\citenamefont {Day}, \citenamefont {Sutula}, \citenamefont {Dietz}, \citenamefont {Raun}, \citenamefont {Sukachev}, \citenamefont {Bhaskar},\ and\ \citenamefont {Hu}}]{day2024electrical}%
  \BibitemOpen
  \bibfield  {author} {\bibinfo {author} {\bibfnamefont {A.~M.}\ \bibnamefont {Day}}, \bibinfo {author} {\bibfnamefont {M.}~\bibnamefont {Sutula}}, \bibinfo {author} {\bibfnamefont {J.~R.}\ \bibnamefont {Dietz}}, \bibinfo {author} {\bibfnamefont {A.}~\bibnamefont {Raun}}, \bibinfo {author} {\bibfnamefont {D.~D.}\ \bibnamefont {Sukachev}}, \bibinfo {author} {\bibfnamefont {M.~K.}\ \bibnamefont {Bhaskar}},\ and\ \bibinfo {author} {\bibfnamefont {E.~L.}\ \bibnamefont {Hu}},\ }\bibfield  {title} {\bibinfo {title} {Electrical manipulation of telecom color centers in silicon},\ }\href@noop {} {\bibfield  {journal} {\bibinfo  {journal} {Nature Communications}\ }\textbf {\bibinfo {volume} {15}},\ \bibinfo {pages} {4722} (\bibinfo {year} {2024})}\BibitemShut {NoStop}%
\bibitem [{\citenamefont {Knall}\ \emph {et~al.}(2022)\citenamefont {Knall}, \citenamefont {Knaut}, \citenamefont {Bekenstein}, \citenamefont {Assumpcao}, \citenamefont {Stroganov}, \citenamefont {Gong}, \citenamefont {Huan}, \citenamefont {Stas}, \citenamefont {Machielse}, \citenamefont {Chalupnik} \emph {et~al.}}]{knall2022efficient}%
  \BibitemOpen
  \bibfield  {author} {\bibinfo {author} {\bibfnamefont {E.~N.}\ \bibnamefont {Knall}}, \bibinfo {author} {\bibfnamefont {C.~M.}\ \bibnamefont {Knaut}}, \bibinfo {author} {\bibfnamefont {R.}~\bibnamefont {Bekenstein}}, \bibinfo {author} {\bibfnamefont {D.~R.}\ \bibnamefont {Assumpcao}}, \bibinfo {author} {\bibfnamefont {P.~L.}\ \bibnamefont {Stroganov}}, \bibinfo {author} {\bibfnamefont {W.}~\bibnamefont {Gong}}, \bibinfo {author} {\bibfnamefont {Y.~Q.}\ \bibnamefont {Huan}}, \bibinfo {author} {\bibfnamefont {P.-J.}\ \bibnamefont {Stas}}, \bibinfo {author} {\bibfnamefont {B.}~\bibnamefont {Machielse}}, \bibinfo {author} {\bibfnamefont {M.}~\bibnamefont {Chalupnik}}, \emph {et~al.},\ }\bibfield  {title} {\bibinfo {title} {Efficient source of shaped single photons based on an integrated diamond nanophotonic system},\ }\href@noop {} {\bibfield  {journal} {\bibinfo  {journal} {Physical Review Letters}\ }\textbf {\bibinfo {volume} {129}},\ \bibinfo {pages} {053603} (\bibinfo {year} {2022})}\BibitemShut {NoStop}%
\bibitem [{\citenamefont {Wang}\ \emph {et~al.}(2021)\citenamefont {Wang}, \citenamefont {Fang}, \citenamefont {Yi}, \citenamefont {Yang}, \citenamefont {Wang}, \citenamefont {Zhou}, \citenamefont {Shen}, \citenamefont {Zhu}, \citenamefont {Zhou}, \citenamefont {Bao} \emph {et~al.}}]{wang2021high}%
  \BibitemOpen
  \bibfield  {author} {\bibinfo {author} {\bibfnamefont {C.}~\bibnamefont {Wang}}, \bibinfo {author} {\bibfnamefont {Z.}~\bibnamefont {Fang}}, \bibinfo {author} {\bibfnamefont {A.}~\bibnamefont {Yi}}, \bibinfo {author} {\bibfnamefont {B.}~\bibnamefont {Yang}}, \bibinfo {author} {\bibfnamefont {Z.}~\bibnamefont {Wang}}, \bibinfo {author} {\bibfnamefont {L.}~\bibnamefont {Zhou}}, \bibinfo {author} {\bibfnamefont {C.}~\bibnamefont {Shen}}, \bibinfo {author} {\bibfnamefont {Y.}~\bibnamefont {Zhu}}, \bibinfo {author} {\bibfnamefont {Y.}~\bibnamefont {Zhou}}, \bibinfo {author} {\bibfnamefont {R.}~\bibnamefont {Bao}}, \emph {et~al.},\ }\bibfield  {title} {\bibinfo {title} {High-q microresonators on 4h-silicon-carbide-on-insulator platform for nonlinear photonics},\ }\href@noop {} {\bibfield  {journal} {\bibinfo  {journal} {Light: Science \& Applications}\ }\textbf {\bibinfo {volume} {10}},\ \bibinfo {pages} {139} (\bibinfo {year} {2021})}\BibitemShut {NoStop}%
\bibitem [{\citenamefont {Assumpcao}\ \emph {et~al.}(2023)\citenamefont {Assumpcao}, \citenamefont {Jin}, \citenamefont {Sutula}, \citenamefont {Ding}, \citenamefont {Pham}, \citenamefont {Knaut}, \citenamefont {Bhaskar}, \citenamefont {Panday}, \citenamefont {Day}, \citenamefont {Renaud} \emph {et~al.}}]{assumpcao2023deterministic}%
  \BibitemOpen
  \bibfield  {author} {\bibinfo {author} {\bibfnamefont {D.~R.}\ \bibnamefont {Assumpcao}}, \bibinfo {author} {\bibfnamefont {C.}~\bibnamefont {Jin}}, \bibinfo {author} {\bibfnamefont {M.}~\bibnamefont {Sutula}}, \bibinfo {author} {\bibfnamefont {S.~W.}\ \bibnamefont {Ding}}, \bibinfo {author} {\bibfnamefont {P.}~\bibnamefont {Pham}}, \bibinfo {author} {\bibfnamefont {C.~M.}\ \bibnamefont {Knaut}}, \bibinfo {author} {\bibfnamefont {M.~K.}\ \bibnamefont {Bhaskar}}, \bibinfo {author} {\bibfnamefont {A.}~\bibnamefont {Panday}}, \bibinfo {author} {\bibfnamefont {A.~M.}\ \bibnamefont {Day}}, \bibinfo {author} {\bibfnamefont {D.}~\bibnamefont {Renaud}}, \emph {et~al.},\ }\bibfield  {title} {\bibinfo {title} {Deterministic creation of strained color centers in nanostructures via high-stress thin films},\ }\href@noop {} {\bibfield  {journal} {\bibinfo  {journal} {Applied Physics Letters}\ }\textbf {\bibinfo {volume} {123}} (\bibinfo {year} {2023})}\BibitemShut {NoStop}%
\bibitem [{\citenamefont {Koller}\ \emph {et~al.}(2025)\citenamefont {Koller}, \citenamefont {Astner}, \citenamefont {Tissot}, \citenamefont {Burkard},\ and\ \citenamefont {Trupke}}]{koller2025strain}%
  \BibitemOpen
  \bibfield  {author} {\bibinfo {author} {\bibfnamefont {P.}~\bibnamefont {Koller}}, \bibinfo {author} {\bibfnamefont {T.}~\bibnamefont {Astner}}, \bibinfo {author} {\bibfnamefont {B.}~\bibnamefont {Tissot}}, \bibinfo {author} {\bibfnamefont {G.}~\bibnamefont {Burkard}},\ and\ \bibinfo {author} {\bibfnamefont {M.}~\bibnamefont {Trupke}},\ }\bibfield  {title} {\bibinfo {title} {Strain-enabled control of the vanadium qudit in silicon carbide},\ }\href@noop {} {\bibfield  {journal} {\bibinfo  {journal} {Physical Review Materials}\ }\textbf {\bibinfo {volume} {9}},\ \bibinfo {pages} {L043201} (\bibinfo {year} {2025})}\BibitemShut {NoStop}%
\bibitem [{\citenamefont {Meesala}\ \emph {et~al.}(2018)\citenamefont {Meesala}, \citenamefont {Sohn}, \citenamefont {Pingault}, \citenamefont {Shao}, \citenamefont {Atikian}, \citenamefont {Holzgrafe}, \citenamefont {Gundogan}, \citenamefont {Stavrakas}, \citenamefont {Sipahigil}, \citenamefont {Chia} \emph {et~al.}}]{meesala2018strain}%
  \BibitemOpen
  \bibfield  {author} {\bibinfo {author} {\bibfnamefont {S.}~\bibnamefont {Meesala}}, \bibinfo {author} {\bibfnamefont {Y.-I.}\ \bibnamefont {Sohn}}, \bibinfo {author} {\bibfnamefont {B.}~\bibnamefont {Pingault}}, \bibinfo {author} {\bibfnamefont {L.}~\bibnamefont {Shao}}, \bibinfo {author} {\bibfnamefont {H.~A.}\ \bibnamefont {Atikian}}, \bibinfo {author} {\bibfnamefont {J.}~\bibnamefont {Holzgrafe}}, \bibinfo {author} {\bibfnamefont {M.}~\bibnamefont {Gundogan}}, \bibinfo {author} {\bibfnamefont {C.}~\bibnamefont {Stavrakas}}, \bibinfo {author} {\bibfnamefont {A.}~\bibnamefont {Sipahigil}}, \bibinfo {author} {\bibfnamefont {C.}~\bibnamefont {Chia}}, \emph {et~al.},\ }\bibfield  {title} {\bibinfo {title} {Strain engineering of the silicon-vacancy center in diamond},\ }\href@noop {} {\bibfield  {journal} {\bibinfo  {journal} {Physical Review B}\ }\textbf {\bibinfo {volume} {97}},\ \bibinfo {pages} {205444} (\bibinfo {year} {2018})}\BibitemShut {NoStop}%
\bibitem [{\citenamefont {Celler}\ and\ \citenamefont {Cristoloveanu}(2003)}]{celler2003frontiers}%
  \BibitemOpen
  \bibfield  {author} {\bibinfo {author} {\bibfnamefont {G.~K.}\ \bibnamefont {Celler}}\ and\ \bibinfo {author} {\bibfnamefont {S.}~\bibnamefont {Cristoloveanu}},\ }\bibfield  {title} {\bibinfo {title} {Frontiers of silicon-on-insulator},\ }\href@noop {} {\bibfield  {journal} {\bibinfo  {journal} {Journal of Applied Physics}\ }\textbf {\bibinfo {volume} {93}},\ \bibinfo {pages} {4955} (\bibinfo {year} {2003})}\BibitemShut {NoStop}%
\bibitem [{\citenamefont {Guo}\ \emph {et~al.}(2021)\citenamefont {Guo}, \citenamefont {Delegan}, \citenamefont {Karsch}, \citenamefont {Li}, \citenamefont {Liu}, \citenamefont {Shreiner}, \citenamefont {Butcher}, \citenamefont {Awschalom}, \citenamefont {Heremans},\ and\ \citenamefont {High}}]{guo2021tunable}%
  \BibitemOpen
  \bibfield  {author} {\bibinfo {author} {\bibfnamefont {X.}~\bibnamefont {Guo}}, \bibinfo {author} {\bibfnamefont {N.}~\bibnamefont {Delegan}}, \bibinfo {author} {\bibfnamefont {J.~C.}\ \bibnamefont {Karsch}}, \bibinfo {author} {\bibfnamefont {Z.}~\bibnamefont {Li}}, \bibinfo {author} {\bibfnamefont {T.}~\bibnamefont {Liu}}, \bibinfo {author} {\bibfnamefont {R.}~\bibnamefont {Shreiner}}, \bibinfo {author} {\bibfnamefont {A.}~\bibnamefont {Butcher}}, \bibinfo {author} {\bibfnamefont {D.~D.}\ \bibnamefont {Awschalom}}, \bibinfo {author} {\bibfnamefont {F.~J.}\ \bibnamefont {Heremans}},\ and\ \bibinfo {author} {\bibfnamefont {A.~A.}\ \bibnamefont {High}},\ }\bibfield  {title} {\bibinfo {title} {Tunable and transferable diamond membranes for integrated quantum technologies},\ }\href@noop {} {\bibfield  {journal} {\bibinfo  {journal} {Nano Letters}\ }\textbf {\bibinfo {volume} {21}},\ \bibinfo {pages} {10392} (\bibinfo {year} {2021})}\BibitemShut {NoStop}%
\bibitem [{\citenamefont {Kimoto}\ and\ \citenamefont {Cooper}(2014)}]{kimoto2014fundamentals}%
  \BibitemOpen
  \bibfield  {author} {\bibinfo {author} {\bibfnamefont {T.}~\bibnamefont {Kimoto}}\ and\ \bibinfo {author} {\bibfnamefont {J.~A.}\ \bibnamefont {Cooper}},\ }\href@noop {} {\emph {\bibinfo {title} {Fundamentals of silicon carbide technology: growth, characterization, devices and applications}}}\ (\bibinfo  {publisher} {John Wiley \& Sons},\ \bibinfo {year} {2014})\BibitemShut {NoStop}%
\bibitem [{\citenamefont {Lukin}\ \emph {et~al.}(2020)\citenamefont {Lukin}, \citenamefont {Dory}, \citenamefont {Guidry}, \citenamefont {Yang}, \citenamefont {Mishra}, \citenamefont {Trivedi}, \citenamefont {Radulaski}, \citenamefont {Sun}, \citenamefont {Vercruysse}, \citenamefont {Ahn} \emph {et~al.}}]{lukin20204h}%
  \BibitemOpen
  \bibfield  {author} {\bibinfo {author} {\bibfnamefont {D.~M.}\ \bibnamefont {Lukin}}, \bibinfo {author} {\bibfnamefont {C.}~\bibnamefont {Dory}}, \bibinfo {author} {\bibfnamefont {M.~A.}\ \bibnamefont {Guidry}}, \bibinfo {author} {\bibfnamefont {K.~Y.}\ \bibnamefont {Yang}}, \bibinfo {author} {\bibfnamefont {S.~D.}\ \bibnamefont {Mishra}}, \bibinfo {author} {\bibfnamefont {R.}~\bibnamefont {Trivedi}}, \bibinfo {author} {\bibfnamefont {M.}~\bibnamefont {Radulaski}}, \bibinfo {author} {\bibfnamefont {S.}~\bibnamefont {Sun}}, \bibinfo {author} {\bibfnamefont {D.}~\bibnamefont {Vercruysse}}, \bibinfo {author} {\bibfnamefont {G.~H.}\ \bibnamefont {Ahn}}, \emph {et~al.},\ }\bibfield  {title} {\bibinfo {title} {4h-silicon-carbide-on-insulator for integrated quantum and nonlinear photonics},\ }\href@noop {} {\bibfield  {journal} {\bibinfo  {journal} {Nature Photonics}\ }\textbf {\bibinfo {volume} {14}},\ \bibinfo {pages} {330} (\bibinfo {year} {2020})}\BibitemShut {NoStop}%
\bibitem [{\citenamefont {Majety}\ \emph {et~al.}(2025)\citenamefont {Majety}, \citenamefont {Norman}, \citenamefont {Saha}, \citenamefont {Rubin}, \citenamefont {Dhuey},\ and\ \citenamefont {Radulaski}}]{majety2025wafer}%
  \BibitemOpen
  \bibfield  {author} {\bibinfo {author} {\bibfnamefont {S.}~\bibnamefont {Majety}}, \bibinfo {author} {\bibfnamefont {V.~A.}\ \bibnamefont {Norman}}, \bibinfo {author} {\bibfnamefont {P.}~\bibnamefont {Saha}}, \bibinfo {author} {\bibfnamefont {A.~H.}\ \bibnamefont {Rubin}}, \bibinfo {author} {\bibfnamefont {S.}~\bibnamefont {Dhuey}},\ and\ \bibinfo {author} {\bibfnamefont {M.}~\bibnamefont {Radulaski}},\ }\bibfield  {title} {\bibinfo {title} {Wafer-scale integration of freestanding photonic devices with color centers in silicon carbide},\ }\href@noop {} {\bibfield  {journal} {\bibinfo  {journal} {npj Nanophotonics}\ }\textbf {\bibinfo {volume} {2}},\ \bibinfo {pages} {3} (\bibinfo {year} {2025})}\BibitemShut {NoStop}%
\bibitem [{\citenamefont {Song}\ \emph {et~al.}(2018)\citenamefont {Song}, \citenamefont {Jeon}, \citenamefont {Kim}, \citenamefont {Kang}, \citenamefont {Asano},\ and\ \citenamefont {Noda}}]{song2018high}%
  \BibitemOpen
  \bibfield  {author} {\bibinfo {author} {\bibfnamefont {B.-S.}\ \bibnamefont {Song}}, \bibinfo {author} {\bibfnamefont {S.}~\bibnamefont {Jeon}}, \bibinfo {author} {\bibfnamefont {H.}~\bibnamefont {Kim}}, \bibinfo {author} {\bibfnamefont {D.~D.}\ \bibnamefont {Kang}}, \bibinfo {author} {\bibfnamefont {T.}~\bibnamefont {Asano}},\ and\ \bibinfo {author} {\bibfnamefont {S.}~\bibnamefont {Noda}},\ }\bibfield  {title} {\bibinfo {title} {High-q-factor nanobeam photonic crystal cavities in bulk silicon carbide},\ }\href@noop {} {\bibfield  {journal} {\bibinfo  {journal} {Applied Physics Letters}\ }\textbf {\bibinfo {volume} {113}} (\bibinfo {year} {2018})}\BibitemShut {NoStop}%
\bibitem [{\citenamefont {Bracher}\ and\ \citenamefont {Hu}(2015)}]{bracher2015fabrication}%
  \BibitemOpen
  \bibfield  {author} {\bibinfo {author} {\bibfnamefont {D.~O.}\ \bibnamefont {Bracher}}\ and\ \bibinfo {author} {\bibfnamefont {E.~L.}\ \bibnamefont {Hu}},\ }\bibfield  {title} {\bibinfo {title} {Fabrication of high-q nanobeam photonic crystals in epitaxially grown 4h-sic},\ }\href@noop {} {\bibfield  {journal} {\bibinfo  {journal} {Nano letters}\ }\textbf {\bibinfo {volume} {15}},\ \bibinfo {pages} {6202} (\bibinfo {year} {2015})}\BibitemShut {NoStop}%
\bibitem [{\citenamefont {Bracher}\ \emph {et~al.}(2017)\citenamefont {Bracher}, \citenamefont {Zhang},\ and\ \citenamefont {Hu}}]{bracher2017selective}%
  \BibitemOpen
  \bibfield  {author} {\bibinfo {author} {\bibfnamefont {D.~O.}\ \bibnamefont {Bracher}}, \bibinfo {author} {\bibfnamefont {X.}~\bibnamefont {Zhang}},\ and\ \bibinfo {author} {\bibfnamefont {E.~L.}\ \bibnamefont {Hu}},\ }\bibfield  {title} {\bibinfo {title} {Selective purcell enhancement of two closely linked zero-phonon transitions of a silicon carbide color center},\ }\href@noop {} {\bibfield  {journal} {\bibinfo  {journal} {Proceedings of the National Academy of Sciences}\ }\textbf {\bibinfo {volume} {114}},\ \bibinfo {pages} {4060} (\bibinfo {year} {2017})}\BibitemShut {NoStop}%
\bibitem [{\citenamefont {Crook}\ \emph {et~al.}(2020)\citenamefont {Crook}, \citenamefont {Anderson}, \citenamefont {Miao}, \citenamefont {Bourassa}, \citenamefont {Lee}, \citenamefont {Bayliss}, \citenamefont {Bracher}, \citenamefont {Zhang}, \citenamefont {Abe}, \citenamefont {Ohshima} \emph {et~al.}}]{crook2020purcell}%
  \BibitemOpen
  \bibfield  {author} {\bibinfo {author} {\bibfnamefont {A.~L.}\ \bibnamefont {Crook}}, \bibinfo {author} {\bibfnamefont {C.~P.}\ \bibnamefont {Anderson}}, \bibinfo {author} {\bibfnamefont {K.~C.}\ \bibnamefont {Miao}}, \bibinfo {author} {\bibfnamefont {A.}~\bibnamefont {Bourassa}}, \bibinfo {author} {\bibfnamefont {H.}~\bibnamefont {Lee}}, \bibinfo {author} {\bibfnamefont {S.~L.}\ \bibnamefont {Bayliss}}, \bibinfo {author} {\bibfnamefont {D.~O.}\ \bibnamefont {Bracher}}, \bibinfo {author} {\bibfnamefont {X.}~\bibnamefont {Zhang}}, \bibinfo {author} {\bibfnamefont {H.}~\bibnamefont {Abe}}, \bibinfo {author} {\bibfnamefont {T.}~\bibnamefont {Ohshima}}, \emph {et~al.},\ }\bibfield  {title} {\bibinfo {title} {Purcell enhancement of a single silicon carbide color center with coherent spin control},\ }\href@noop {} {\bibfield  {journal} {\bibinfo  {journal} {Nano letters}\ }\textbf {\bibinfo {volume} {20}},\ \bibinfo {pages} {3427} (\bibinfo {year} {2020})}\BibitemShut {NoStop}%
\bibitem [{\citenamefont {Tsukimoto}\ \emph {et~al.}(2018)\citenamefont {Tsukimoto}, \citenamefont {Ise}, \citenamefont {Maruyama}, \citenamefont {Hashimoto}, \citenamefont {Sakurada}, \citenamefont {Senzaki}, \citenamefont {Kato}, \citenamefont {Kojima},\ and\ \citenamefont {Okumura}}]{tsukimoto2018local}%
  \BibitemOpen
  \bibfield  {author} {\bibinfo {author} {\bibfnamefont {S.}~\bibnamefont {Tsukimoto}}, \bibinfo {author} {\bibfnamefont {T.}~\bibnamefont {Ise}}, \bibinfo {author} {\bibfnamefont {G.}~\bibnamefont {Maruyama}}, \bibinfo {author} {\bibfnamefont {S.}~\bibnamefont {Hashimoto}}, \bibinfo {author} {\bibfnamefont {T.}~\bibnamefont {Sakurada}}, \bibinfo {author} {\bibfnamefont {J.}~\bibnamefont {Senzaki}}, \bibinfo {author} {\bibfnamefont {T.}~\bibnamefont {Kato}}, \bibinfo {author} {\bibfnamefont {K.}~\bibnamefont {Kojima}},\ and\ \bibinfo {author} {\bibfnamefont {H.}~\bibnamefont {Okumura}},\ }\bibfield  {title} {\bibinfo {title} {Local strain distribution and microstructure of grinding-induced damage layers in sic wafer},\ }\href@noop {} {\bibfield  {journal} {\bibinfo  {journal} {Journal of Electronic Materials}\ }\textbf {\bibinfo {volume} {47}},\ \bibinfo {pages} {6722} (\bibinfo {year} {2018})}\BibitemShut {NoStop}%
\bibitem [{\citenamefont {Kozlov}\ \emph {et~al.}(2002)\citenamefont {Kozlov}, \citenamefont {Kozlovskii}, \citenamefont {Titkov}, \citenamefont {Dunaevskii},\ and\ \citenamefont {Kryzhanovskii}}]{kozlov2002buried}%
  \BibitemOpen
  \bibfield  {author} {\bibinfo {author} {\bibfnamefont {V.}~\bibnamefont {Kozlov}}, \bibinfo {author} {\bibfnamefont {V.}~\bibnamefont {Kozlovskii}}, \bibinfo {author} {\bibfnamefont {A.}~\bibnamefont {Titkov}}, \bibinfo {author} {\bibfnamefont {M.}~\bibnamefont {Dunaevskii}},\ and\ \bibinfo {author} {\bibfnamefont {A.}~\bibnamefont {Kryzhanovskii}},\ }\bibfield  {title} {\bibinfo {title} {Buried nanoscale damaged layers formed in si and sic crystals as a result of high-dose proton implantation},\ }\href@noop {} {\bibfield  {journal} {\bibinfo  {journal} {Semiconductors}\ }\textbf {\bibinfo {volume} {36}},\ \bibinfo {pages} {1227} (\bibinfo {year} {2002})}\BibitemShut {NoStop}%
\bibitem [{\citenamefont {Dietz}\ \emph {et~al.}(2025)\citenamefont {Dietz}, \citenamefont {Xie}, \citenamefont {Day},\ and\ \citenamefont {Hu}}]{dietz2025selective}%
  \BibitemOpen
  \bibfield  {author} {\bibinfo {author} {\bibfnamefont {J.}~\bibnamefont {Dietz}}, \bibinfo {author} {\bibfnamefont {A.}~\bibnamefont {Xie}}, \bibinfo {author} {\bibfnamefont {A.~M.}\ \bibnamefont {Day}},\ and\ \bibinfo {author} {\bibfnamefont {E.~L.}\ \bibnamefont {Hu}},\ }\bibfield  {title} {\bibinfo {title} {Selective undercut of undoped optical membranes for spin-active color centers in 4h-silicon carbide},\ }\href@noop {} {\bibfield  {journal} {\bibinfo  {journal} {ACS nano}\ } (\bibinfo {year} {2025})}\BibitemShut {NoStop}%
\bibitem [{\citenamefont {Dietz}(2025)}]{dietz2025integrating}%
  \BibitemOpen
  \bibfield  {author} {\bibinfo {author} {\bibfnamefont {J.~R.}\ \bibnamefont {Dietz}},\ }\emph {\bibinfo {title} {Integrating Spin Defects With Thin-Film Silicon Carbide Devices}},\ \href@noop {} {Ph.D. thesis},\ \bibinfo  {school} {Harvard University} (\bibinfo {year} {2025})\BibitemShut {NoStop}%
\bibitem [{\citenamefont {Zhang}\ and\ \citenamefont {Yates~Jr}(2012)}]{zhang2012band}%
  \BibitemOpen
  \bibfield  {author} {\bibinfo {author} {\bibfnamefont {Z.}~\bibnamefont {Zhang}}\ and\ \bibinfo {author} {\bibfnamefont {J.~T.}\ \bibnamefont {Yates~Jr}},\ }\bibfield  {title} {\bibinfo {title} {Band bending in semiconductors: chemical and physical consequences at surfaces and interfaces},\ }\href@noop {} {\bibfield  {journal} {\bibinfo  {journal} {Chemical reviews}\ }\textbf {\bibinfo {volume} {112}},\ \bibinfo {pages} {5520} (\bibinfo {year} {2012})}\BibitemShut {NoStop}%
\bibitem [{\citenamefont {Pavunny}\ \emph {et~al.}(2019)\citenamefont {Pavunny}, \citenamefont {Myers-Ward}, \citenamefont {Daniels}, \citenamefont {Shi}, \citenamefont {Sridhara}, \citenamefont {DeJarld}, \citenamefont {Boyd}, \citenamefont {Kub}, \citenamefont {Kohl}, \citenamefont {Carter} \emph {et~al.}}]{pavunny2019doping}%
  \BibitemOpen
  \bibfield  {author} {\bibinfo {author} {\bibfnamefont {S.~P.}\ \bibnamefont {Pavunny}}, \bibinfo {author} {\bibfnamefont {R.~L.}\ \bibnamefont {Myers-Ward}}, \bibinfo {author} {\bibfnamefont {K.~M.}\ \bibnamefont {Daniels}}, \bibinfo {author} {\bibfnamefont {W.}~\bibnamefont {Shi}}, \bibinfo {author} {\bibfnamefont {K.}~\bibnamefont {Sridhara}}, \bibinfo {author} {\bibfnamefont {M.~T.}\ \bibnamefont {DeJarld}}, \bibinfo {author} {\bibfnamefont {A.~K.}\ \bibnamefont {Boyd}}, \bibinfo {author} {\bibfnamefont {F.~J.}\ \bibnamefont {Kub}}, \bibinfo {author} {\bibfnamefont {P.~A.}\ \bibnamefont {Kohl}}, \bibinfo {author} {\bibfnamefont {S.~G.}\ \bibnamefont {Carter}}, \emph {et~al.},\ }\bibfield  {title} {\bibinfo {title} {On the doping concentration dependence and dopant selectivity of photogenerated carrier assisted etching of 4h--sic epilayers},\ }\href@noop {} {\bibfield  {journal} {\bibinfo  {journal} {Electrochimica Acta}\ }\textbf {\bibinfo {volume} {323}},\ \bibinfo {pages} {134778} (\bibinfo {year}
  {2019})}\BibitemShut {NoStop}%
\bibitem [{\citenamefont {Giovannetti}\ \emph {et~al.}(2011)\citenamefont {Giovannetti}, \citenamefont {Lloyd},\ and\ \citenamefont {Maccone}}]{giovannetti2011advances}%
  \BibitemOpen
  \bibfield  {author} {\bibinfo {author} {\bibfnamefont {V.}~\bibnamefont {Giovannetti}}, \bibinfo {author} {\bibfnamefont {S.}~\bibnamefont {Lloyd}},\ and\ \bibinfo {author} {\bibfnamefont {L.}~\bibnamefont {Maccone}},\ }\bibfield  {title} {\bibinfo {title} {Advances in quantum metrology},\ }\href@noop {} {\bibfield  {journal} {\bibinfo  {journal} {Nature photonics}\ }\textbf {\bibinfo {volume} {5}},\ \bibinfo {pages} {222} (\bibinfo {year} {2011})}\BibitemShut {NoStop}%
\bibitem [{\citenamefont {Xie}(2025)}]{SiCNBcode}%
  \BibitemOpen
  \bibfield  {author} {\bibinfo {author} {\bibfnamefont {A.}~\bibnamefont {Xie}},\ }\href {https://www.flexcompute.com/tidy3d/kjFIWnQUtc} {\bibinfo {title} {{4H-SiC 1D Photonic Crystal Cavities}}} (\bibinfo {year} {2025})\BibitemShut {NoStop}%
\bibitem [{\citenamefont {Day}\ \emph {et~al.}(2023)\citenamefont {Day}, \citenamefont {Dietz}, \citenamefont {Sutula}, \citenamefont {Yeh},\ and\ \citenamefont {Hu}}]{day2023laser}%
  \BibitemOpen
  \bibfield  {author} {\bibinfo {author} {\bibfnamefont {A.~M.}\ \bibnamefont {Day}}, \bibinfo {author} {\bibfnamefont {J.~R.}\ \bibnamefont {Dietz}}, \bibinfo {author} {\bibfnamefont {M.}~\bibnamefont {Sutula}}, \bibinfo {author} {\bibfnamefont {M.}~\bibnamefont {Yeh}},\ and\ \bibinfo {author} {\bibfnamefont {E.~L.}\ \bibnamefont {Hu}},\ }\bibfield  {title} {\bibinfo {title} {Laser writing of spin defects in nanophotonic cavities},\ }\href@noop {} {\bibfield  {journal} {\bibinfo  {journal} {Nature Materials}\ }\textbf {\bibinfo {volume} {22}},\ \bibinfo {pages} {696} (\bibinfo {year} {2023})}\BibitemShut {NoStop}%
\bibitem [{\citenamefont {Jin}(2025)}]{SiCTCcode}%
  \BibitemOpen
  \bibfield  {author} {\bibinfo {author} {\bibfnamefont {C.}~\bibnamefont {Jin}},\ }\href {https://www.flexcompute.com/tidy3d/fggro9Y3gb} {\bibinfo {title} {{4H-SiC 1D Photonic Crystal Cavities With Tapered Fiber Interface}}} (\bibinfo {year} {2025})\BibitemShut {NoStop}%
\bibitem [{\citenamefont {Bersin}\ \emph {et~al.}(2024{\natexlab{b}})\citenamefont {Bersin}, \citenamefont {Grein}, \citenamefont {Sutula}, \citenamefont {Murphy}, \citenamefont {Huan}, \citenamefont {Stevens}, \citenamefont {Suleymanzade}, \citenamefont {Lee}, \citenamefont {Riedinger}, \citenamefont {Starling} \emph {et~al.}}]{bersin2024development}%
  \BibitemOpen
  \bibfield  {author} {\bibinfo {author} {\bibfnamefont {E.}~\bibnamefont {Bersin}}, \bibinfo {author} {\bibfnamefont {M.}~\bibnamefont {Grein}}, \bibinfo {author} {\bibfnamefont {M.}~\bibnamefont {Sutula}}, \bibinfo {author} {\bibfnamefont {R.}~\bibnamefont {Murphy}}, \bibinfo {author} {\bibfnamefont {Y.~Q.}\ \bibnamefont {Huan}}, \bibinfo {author} {\bibfnamefont {M.}~\bibnamefont {Stevens}}, \bibinfo {author} {\bibfnamefont {A.}~\bibnamefont {Suleymanzade}}, \bibinfo {author} {\bibfnamefont {C.}~\bibnamefont {Lee}}, \bibinfo {author} {\bibfnamefont {R.}~\bibnamefont {Riedinger}}, \bibinfo {author} {\bibfnamefont {D.~J.}\ \bibnamefont {Starling}}, \emph {et~al.},\ }\bibfield  {title} {\bibinfo {title} {Development of a boston-area 50-km fiber quantum network testbed},\ }\href@noop {} {\bibfield  {journal} {\bibinfo  {journal} {Physical Review Applied}\ }\textbf {\bibinfo {volume} {21}},\ \bibinfo {pages} {014024} (\bibinfo {year} {2024}{\natexlab{b}})}\BibitemShut {NoStop}%
\bibitem [{\citenamefont {Vasselon}\ \emph {et~al.}(2023)\citenamefont {Vasselon}, \citenamefont {Hern{\'a}ndez-M{\'\i}nguez}, \citenamefont {Hollenbach}, \citenamefont {Astakhov},\ and\ \citenamefont {Santos}}]{vasselon2023acoustically}%
  \BibitemOpen
  \bibfield  {author} {\bibinfo {author} {\bibfnamefont {T.}~\bibnamefont {Vasselon}}, \bibinfo {author} {\bibfnamefont {A.}~\bibnamefont {Hern{\'a}ndez-M{\'\i}nguez}}, \bibinfo {author} {\bibfnamefont {M.}~\bibnamefont {Hollenbach}}, \bibinfo {author} {\bibfnamefont {G.}~\bibnamefont {Astakhov}},\ and\ \bibinfo {author} {\bibfnamefont {P.}~\bibnamefont {Santos}},\ }\bibfield  {title} {\bibinfo {title} {Acoustically induced spin resonances of silicon-vacancy centers in 4 h-sic},\ }\href@noop {} {\bibfield  {journal} {\bibinfo  {journal} {Physical Review Applied}\ }\textbf {\bibinfo {volume} {20}},\ \bibinfo {pages} {034017} (\bibinfo {year} {2023})}\BibitemShut {NoStop}%
\bibitem [{\citenamefont {Schutz}\ and\ \citenamefont {Schutz}(2017)}]{schutz2017universal}%
  \BibitemOpen
  \bibfield  {author} {\bibinfo {author} {\bibfnamefont {M.~J.}\ \bibnamefont {Schutz}}\ and\ \bibinfo {author} {\bibfnamefont {M.~J.}\ \bibnamefont {Schutz}},\ }\bibfield  {title} {\bibinfo {title} {Universal quantum transducers based on surface acoustic waves},\ }\href@noop {} {\bibfield  {journal} {\bibinfo  {journal} {Quantum dots for quantum information processing: controlling and exploiting the quantum dot environment}\ ,\ \bibinfo {pages} {143}} (\bibinfo {year} {2017})}\BibitemShut {NoStop}%
\bibitem [{\citenamefont {Chen}\ \emph {et~al.}(2020)\citenamefont {Chen}, \citenamefont {Bhave},\ and\ \citenamefont {Fuchs}}]{chen2020acoustically}%
  \BibitemOpen
  \bibfield  {author} {\bibinfo {author} {\bibfnamefont {H.}~\bibnamefont {Chen}}, \bibinfo {author} {\bibfnamefont {S.}~\bibnamefont {Bhave}},\ and\ \bibinfo {author} {\bibfnamefont {G.}~\bibnamefont {Fuchs}},\ }\bibfield  {title} {\bibinfo {title} {Acoustically driving the single-quantum spin transition of diamond nitrogen-vacancy centers},\ }\href@noop {} {\bibfield  {journal} {\bibinfo  {journal} {Physical Review Applied}\ }\textbf {\bibinfo {volume} {13}},\ \bibinfo {pages} {054068} (\bibinfo {year} {2020})}\BibitemShut {NoStop}%
\bibitem [{\citenamefont {Soykal}\ \emph {et~al.}(2016)\citenamefont {Soykal}, \citenamefont {Dev},\ and\ \citenamefont {Economou}}]{soykal2016silicon}%
  \BibitemOpen
  \bibfield  {author} {\bibinfo {author} {\bibfnamefont {O.}~\bibnamefont {Soykal}}, \bibinfo {author} {\bibfnamefont {P.}~\bibnamefont {Dev}},\ and\ \bibinfo {author} {\bibfnamefont {S.~E.}\ \bibnamefont {Economou}},\ }\bibfield  {title} {\bibinfo {title} {Silicon vacancy center in 4 h-sic: Electronic structure and spin-photon interfaces},\ }\href@noop {} {\bibfield  {journal} {\bibinfo  {journal} {Physical Review B}\ }\textbf {\bibinfo {volume} {93}},\ \bibinfo {pages} {081207} (\bibinfo {year} {2016})}\BibitemShut {NoStop}%
\bibitem [{\citenamefont {Olsson~III}\ \emph {et~al.}(2014)\citenamefont {Olsson~III}, \citenamefont {Hattar}, \citenamefont {Homeijer}, \citenamefont {Wiwi}, \citenamefont {Eichenfield}, \citenamefont {Branch}, \citenamefont {Baker}, \citenamefont {Nguyen}, \citenamefont {Clark}, \citenamefont {Bauer} \emph {et~al.}}]{olsson2014high}%
  \BibitemOpen
  \bibfield  {author} {\bibinfo {author} {\bibfnamefont {R.~H.}\ \bibnamefont {Olsson~III}}, \bibinfo {author} {\bibfnamefont {K.}~\bibnamefont {Hattar}}, \bibinfo {author} {\bibfnamefont {S.~J.}\ \bibnamefont {Homeijer}}, \bibinfo {author} {\bibfnamefont {M.}~\bibnamefont {Wiwi}}, \bibinfo {author} {\bibfnamefont {M.}~\bibnamefont {Eichenfield}}, \bibinfo {author} {\bibfnamefont {D.~W.}\ \bibnamefont {Branch}}, \bibinfo {author} {\bibfnamefont {M.~S.}\ \bibnamefont {Baker}}, \bibinfo {author} {\bibfnamefont {J.}~\bibnamefont {Nguyen}}, \bibinfo {author} {\bibfnamefont {B.}~\bibnamefont {Clark}}, \bibinfo {author} {\bibfnamefont {T.}~\bibnamefont {Bauer}}, \emph {et~al.},\ }\bibfield  {title} {\bibinfo {title} {A high electromechanical coupling coefficient sh0 lamb wave lithium niobate micromechanical resonator and a method for fabrication},\ }\href@noop {} {\bibfield  {journal} {\bibinfo  {journal} {Sensors and Actuators A: Physical}\ }\textbf {\bibinfo {volume} {209}},\ \bibinfo {pages} {183} (\bibinfo
  {year} {2014})}\BibitemShut {NoStop}%
\bibitem [{\citenamefont {Xu}\ \emph {et~al.}(2025)\citenamefont {Xu}, \citenamefont {Ding}, \citenamefont {Cornell}, \citenamefont {Mohideen}, \citenamefont {Yeh}, \citenamefont {Kuruma}, \citenamefont {Magalhaes}, \citenamefont {Shams-Ansari}, \citenamefont {Pingault},\ and\ \citenamefont {Loncar}}]{xu2025thin}%
  \BibitemOpen
  \bibfield  {author} {\bibinfo {author} {\bibfnamefont {Z.}~\bibnamefont {Xu}}, \bibinfo {author} {\bibfnamefont {S.~W.}\ \bibnamefont {Ding}}, \bibinfo {author} {\bibfnamefont {E.}~\bibnamefont {Cornell}}, \bibinfo {author} {\bibfnamefont {S.}~\bibnamefont {Mohideen}}, \bibinfo {author} {\bibfnamefont {M.}~\bibnamefont {Yeh}}, \bibinfo {author} {\bibfnamefont {K.}~\bibnamefont {Kuruma}}, \bibinfo {author} {\bibfnamefont {L.}~\bibnamefont {Magalhaes}}, \bibinfo {author} {\bibfnamefont {A.}~\bibnamefont {Shams-Ansari}}, \bibinfo {author} {\bibfnamefont {B.}~\bibnamefont {Pingault}},\ and\ \bibinfo {author} {\bibfnamefont {M.}~\bibnamefont {Loncar}},\ }\bibfield  {title} {\bibinfo {title} {Thin film lithium niobate on diamond (linda) platform for efficient spin-phonon coupling},\ }\href@noop {} {\bibfield  {journal} {\bibinfo  {journal} {arXiv preprint arXiv:2505.08895}\ } (\bibinfo {year} {2025})}\BibitemShut {NoStop}%
\bibitem [{\citenamefont {Li}\ \emph {et~al.}(2020)\citenamefont {Li}, \citenamefont {Ling}, \citenamefont {He}, \citenamefont {Javid}, \citenamefont {Xue},\ and\ \citenamefont {Lin}}]{li2020lithium}%
  \BibitemOpen
  \bibfield  {author} {\bibinfo {author} {\bibfnamefont {M.}~\bibnamefont {Li}}, \bibinfo {author} {\bibfnamefont {J.}~\bibnamefont {Ling}}, \bibinfo {author} {\bibfnamefont {Y.}~\bibnamefont {He}}, \bibinfo {author} {\bibfnamefont {U.~A.}\ \bibnamefont {Javid}}, \bibinfo {author} {\bibfnamefont {S.}~\bibnamefont {Xue}},\ and\ \bibinfo {author} {\bibfnamefont {Q.}~\bibnamefont {Lin}},\ }\bibfield  {title} {\bibinfo {title} {Lithium niobate photonic-crystal electro-optic modulator},\ }\href@noop {} {\bibfield  {journal} {\bibinfo  {journal} {Nature communications}\ }\textbf {\bibinfo {volume} {11}},\ \bibinfo {pages} {4123} (\bibinfo {year} {2020})}\BibitemShut {NoStop}%
\bibitem [{\citenamefont {Xia}\ \emph {et~al.}(2022)\citenamefont {Xia}, \citenamefont {Sardi}, \citenamefont {Sauerzapf}, \citenamefont {Kornher}, \citenamefont {Becker}, \citenamefont {Kis}, \citenamefont {Kovacs}, \citenamefont {Dertli}, \citenamefont {Foglszinger}, \citenamefont {Kolesov} \emph {et~al.}}]{xia2022}%
  \BibitemOpen
  \bibfield  {author} {\bibinfo {author} {\bibfnamefont {K.}~\bibnamefont {Xia}}, \bibinfo {author} {\bibfnamefont {F.}~\bibnamefont {Sardi}}, \bibinfo {author} {\bibfnamefont {C.}~\bibnamefont {Sauerzapf}}, \bibinfo {author} {\bibfnamefont {T.}~\bibnamefont {Kornher}}, \bibinfo {author} {\bibfnamefont {H.-W.}\ \bibnamefont {Becker}}, \bibinfo {author} {\bibfnamefont {Z.}~\bibnamefont {Kis}}, \bibinfo {author} {\bibfnamefont {L.}~\bibnamefont {Kovacs}}, \bibinfo {author} {\bibfnamefont {D.}~\bibnamefont {Dertli}}, \bibinfo {author} {\bibfnamefont {J.}~\bibnamefont {Foglszinger}}, \bibinfo {author} {\bibfnamefont {R.}~\bibnamefont {Kolesov}}, \emph {et~al.},\ }\bibfield  {title} {\bibinfo {title} {Tunable microcavities coupled to rare-earth quantum emitters},\ }\href@noop {} {\bibfield  {journal} {\bibinfo  {journal} {Optica}\ }\textbf {\bibinfo {volume} {9}},\ \bibinfo {pages} {445} (\bibinfo {year} {2022})}\BibitemShut {NoStop}%
\bibitem [{\citenamefont {Assumpcao}\ \emph {et~al.}(2024)\citenamefont {Assumpcao}, \citenamefont {Renaud}, \citenamefont {Baradari}, \citenamefont {Zeng}, \citenamefont {De-Eknamkul}, \citenamefont {Xin}, \citenamefont {Shams-Ansari}, \citenamefont {Barton}, \citenamefont {Machielse},\ and\ \citenamefont {Loncar}}]{assumpcao2024thin}%
  \BibitemOpen
  \bibfield  {author} {\bibinfo {author} {\bibfnamefont {D.}~\bibnamefont {Assumpcao}}, \bibinfo {author} {\bibfnamefont {D.}~\bibnamefont {Renaud}}, \bibinfo {author} {\bibfnamefont {A.}~\bibnamefont {Baradari}}, \bibinfo {author} {\bibfnamefont {B.}~\bibnamefont {Zeng}}, \bibinfo {author} {\bibfnamefont {C.}~\bibnamefont {De-Eknamkul}}, \bibinfo {author} {\bibfnamefont {C.}~\bibnamefont {Xin}}, \bibinfo {author} {\bibfnamefont {A.}~\bibnamefont {Shams-Ansari}}, \bibinfo {author} {\bibfnamefont {D.}~\bibnamefont {Barton}}, \bibinfo {author} {\bibfnamefont {B.}~\bibnamefont {Machielse}},\ and\ \bibinfo {author} {\bibfnamefont {M.}~\bibnamefont {Loncar}},\ }\bibfield  {title} {\bibinfo {title} {A thin film lithium niobate near-infrared platform for multiplexing quantum nodes},\ }\href@noop {} {\bibfield  {journal} {\bibinfo  {journal} {Nature communications}\ }\textbf {\bibinfo {volume} {15}},\ \bibinfo {pages} {1} (\bibinfo {year} {2024})}\BibitemShut {NoStop}%
\bibitem [{\citenamefont {Krumrein}\ \emph {et~al.}(2024)\citenamefont {Krumrein}, \citenamefont {Nold}, \citenamefont {Davidson-Marquis}, \citenamefont {Bouamra}, \citenamefont {Niechziol}, \citenamefont {Steidl}, \citenamefont {Peng}, \citenamefont {Korber}, \citenamefont {Stohr}, \citenamefont {Gross} \emph {et~al.}}]{krumrein2024precise}%
  \BibitemOpen
  \bibfield  {author} {\bibinfo {author} {\bibfnamefont {M.}~\bibnamefont {Krumrein}}, \bibinfo {author} {\bibfnamefont {R.}~\bibnamefont {Nold}}, \bibinfo {author} {\bibfnamefont {F.}~\bibnamefont {Davidson-Marquis}}, \bibinfo {author} {\bibfnamefont {A.}~\bibnamefont {Bouamra}}, \bibinfo {author} {\bibfnamefont {L.}~\bibnamefont {Niechziol}}, \bibinfo {author} {\bibfnamefont {T.}~\bibnamefont {Steidl}}, \bibinfo {author} {\bibfnamefont {R.}~\bibnamefont {Peng}}, \bibinfo {author} {\bibfnamefont {J.}~\bibnamefont {Korber}}, \bibinfo {author} {\bibfnamefont {R.}~\bibnamefont {Stohr}}, \bibinfo {author} {\bibfnamefont {N.}~\bibnamefont {Gross}}, \emph {et~al.},\ }\bibfield  {title} {\bibinfo {title} {Precise characterization of a waveguide fiber interface in silicon carbide},\ }\href@noop {} {\bibfield  {journal} {\bibinfo  {journal} {ACS photonics}\ }\textbf {\bibinfo {volume} {11}},\ \bibinfo {pages} {2160} (\bibinfo {year} {2024})}\BibitemShut {NoStop}%
\bibitem [{\citenamefont {Campbell}\ \emph {et~al.}(2024)\citenamefont {Campbell}, \citenamefont {Anderson}, \citenamefont {Hsu}, \citenamefont {Barrera}, \citenamefont {Kramer}, \citenamefont {Cho}, \citenamefont {Chulukhadze}, \citenamefont {Latham}, \citenamefont {Li},\ and\ \citenamefont {Lu}}]{campbell202421}%
  \BibitemOpen
  \bibfield  {author} {\bibinfo {author} {\bibfnamefont {J.}~\bibnamefont {Campbell}}, \bibinfo {author} {\bibfnamefont {I.}~\bibnamefont {Anderson}}, \bibinfo {author} {\bibfnamefont {T.-H.}\ \bibnamefont {Hsu}}, \bibinfo {author} {\bibfnamefont {O.~A.}\ \bibnamefont {Barrera}}, \bibinfo {author} {\bibfnamefont {J.}~\bibnamefont {Kramer}}, \bibinfo {author} {\bibfnamefont {S.}~\bibnamefont {Cho}}, \bibinfo {author} {\bibfnamefont {V.}~\bibnamefont {Chulukhadze}}, \bibinfo {author} {\bibfnamefont {G.}~\bibnamefont {Latham}}, \bibinfo {author} {\bibfnamefont {M.-H.}\ \bibnamefont {Li}},\ and\ \bibinfo {author} {\bibfnamefont {R.}~\bibnamefont {Lu}},\ }\bibfield  {title} {\bibinfo {title} {21.4 ghz surface acoustic wave resonator with 11,400 m/s phase velocity in thin-film lithium niobate on silicon carbide},\ }in\ \href@noop {} {\emph {\bibinfo {booktitle} {2024 IEEE Ultrasonics, Ferroelectrics, and Frequency Control Joint Symposium (UFFC-JS)}}}\ (\bibinfo {organization} {IEEE},\ \bibinfo {year} {2024})\ pp.\
  \bibinfo {pages} {1--4}\BibitemShut {NoStop}%
\bibitem [{\citenamefont {Hochreiter}\ \emph {et~al.}(2023)\citenamefont {Hochreiter}, \citenamefont {Gross}, \citenamefont {Moller}, \citenamefont {Krieger},\ and\ \citenamefont {Weber}}]{hochreiter2023electrochemical}%
  \BibitemOpen
  \bibfield  {author} {\bibinfo {author} {\bibfnamefont {A.}~\bibnamefont {Hochreiter}}, \bibinfo {author} {\bibfnamefont {F.}~\bibnamefont {Gross}}, \bibinfo {author} {\bibfnamefont {M.-N.}\ \bibnamefont {Moller}}, \bibinfo {author} {\bibfnamefont {M.}~\bibnamefont {Krieger}},\ and\ \bibinfo {author} {\bibfnamefont {H.~B.}\ \bibnamefont {Weber}},\ }\bibfield  {title} {\bibinfo {title} {Electrochemical etching strategy for shaping monolithic 3d structures from 4h-sic wafers},\ }\href@noop {} {\bibfield  {journal} {\bibinfo  {journal} {Scientific reports}\ }\textbf {\bibinfo {volume} {13}},\ \bibinfo {pages} {19086} (\bibinfo {year} {2023})}\BibitemShut {NoStop}%
\bibitem [{\citenamefont {Das}\ \emph {et~al.}(2024)\citenamefont {Das}, \citenamefont {Lichtenwalner}, \citenamefont {Dixit}, \citenamefont {Rogers}, \citenamefont {Scholze},\ and\ \citenamefont {Ryu}}]{das2024study}%
  \BibitemOpen
  \bibfield  {author} {\bibinfo {author} {\bibfnamefont {S.}~\bibnamefont {Das}}, \bibinfo {author} {\bibfnamefont {D.~J.}\ \bibnamefont {Lichtenwalner}}, \bibinfo {author} {\bibfnamefont {H.}~\bibnamefont {Dixit}}, \bibinfo {author} {\bibfnamefont {S.}~\bibnamefont {Rogers}}, \bibinfo {author} {\bibfnamefont {A.}~\bibnamefont {Scholze}},\ and\ \bibinfo {author} {\bibfnamefont {S.-H.}\ \bibnamefont {Ryu}},\ }\bibfield  {title} {\bibinfo {title} {Study of dopant activation and ionization for phosphorus in 4h-sic},\ }\href@noop {} {\bibfield  {journal} {\bibinfo  {journal} {Journal of Electronic Materials}\ }\textbf {\bibinfo {volume} {53}},\ \bibinfo {pages} {2806} (\bibinfo {year} {2024})}\BibitemShut {NoStop}%
\bibitem [{\citenamefont {Negoro}\ \emph {et~al.}(2004)\citenamefont {Negoro}, \citenamefont {Kimoto}, \citenamefont {Matsunami}, \citenamefont {Schmid},\ and\ \citenamefont {Pensl}}]{negoro2004electrical}%
  \BibitemOpen
  \bibfield  {author} {\bibinfo {author} {\bibfnamefont {Y.}~\bibnamefont {Negoro}}, \bibinfo {author} {\bibfnamefont {T.}~\bibnamefont {Kimoto}}, \bibinfo {author} {\bibfnamefont {H.}~\bibnamefont {Matsunami}}, \bibinfo {author} {\bibfnamefont {F.}~\bibnamefont {Schmid}},\ and\ \bibinfo {author} {\bibfnamefont {G.}~\bibnamefont {Pensl}},\ }\bibfield  {title} {\bibinfo {title} {Electrical activation of high-concentration aluminum implanted in 4h-sic},\ }\href@noop {} {\bibfield  {journal} {\bibinfo  {journal} {Journal of Applied Physics}\ }\textbf {\bibinfo {volume} {96}},\ \bibinfo {pages} {4916} (\bibinfo {year} {2004})}\BibitemShut {NoStop}%
\bibitem [{\citenamefont {Wolfowicz}\ \emph {et~al.}(2020)\citenamefont {Wolfowicz}, \citenamefont {Anderson}, \citenamefont {Diler}, \citenamefont {Poluektov}, \citenamefont {Heremans},\ and\ \citenamefont {Awschalom}}]{wolfowicz2020vanadium}%
  \BibitemOpen
  \bibfield  {author} {\bibinfo {author} {\bibfnamefont {G.}~\bibnamefont {Wolfowicz}}, \bibinfo {author} {\bibfnamefont {C.~P.}\ \bibnamefont {Anderson}}, \bibinfo {author} {\bibfnamefont {B.}~\bibnamefont {Diler}}, \bibinfo {author} {\bibfnamefont {O.~G.}\ \bibnamefont {Poluektov}}, \bibinfo {author} {\bibfnamefont {F.~J.}\ \bibnamefont {Heremans}},\ and\ \bibinfo {author} {\bibfnamefont {D.~D.}\ \bibnamefont {Awschalom}},\ }\bibfield  {title} {\bibinfo {title} {Vanadium spin qubits as telecom quantum emitters in silicon carbide},\ }\href@noop {} {\bibfield  {journal} {\bibinfo  {journal} {Science advances}\ }\textbf {\bibinfo {volume} {6}},\ \bibinfo {pages} {eaaz1192} (\bibinfo {year} {2020})}\BibitemShut {NoStop}%
\end{thebibliography}%


\begin{thebibliography}{5}%
\makeatletter
\providecommand \@ifxundefined [1]{%
 \@ifx{#1\undefined}
}%
\providecommand \@ifnum [1]{%
 \ifnum #1\expandafter \@firstoftwo
 \else \expandafter \@secondoftwo
 \fi
}%
\providecommand \@ifx [1]{%
 \ifx #1\expandafter \@firstoftwo
 \else \expandafter \@secondoftwo
 \fi
}%
\providecommand \natexlab [1]{#1}%
\providecommand \enquote  [1]{``#1''}%
\providecommand \bibnamefont  [1]{#1}%
\providecommand \bibfnamefont [1]{#1}%
\providecommand \citenamefont [1]{#1}%
\providecommand \href@noop [0]{\@secondoftwo}%
\providecommand \href [0]{\begingroup \@sanitize@url \@href}%
\providecommand \@href[1]{\@@startlink{#1}\@@href}%
\providecommand \@@href[1]{\endgroup#1\@@endlink}%
\providecommand \@sanitize@url [0]{\catcode `\\12\catcode `\$12\catcode `\&12\catcode `\#12\catcode `\^12\catcode `\_12\catcode `\%12\relax}%
\providecommand \@@startlink[1]{}%
\providecommand \@@endlink[0]{}%
\providecommand \url  [0]{\begingroup\@sanitize@url \@url }%
\providecommand \@url [1]{\endgroup\@href {#1}{\urlprefix }}%
\providecommand \urlprefix  [0]{URL }%
\providecommand \Eprint [0]{\href }%
\providecommand \doibase [0]{https://doi.org/}%
\providecommand \selectlanguage [0]{\@gobble}%
\providecommand \bibinfo  [0]{\@secondoftwo}%
\providecommand \bibfield  [0]{\@secondoftwo}%
\providecommand \translation [1]{[#1]}%
\providecommand \BibitemOpen [0]{}%
\providecommand \bibitemStop [0]{}%
\providecommand \bibitemNoStop [0]{.\EOS\space}%
\providecommand \EOS [0]{\spacefactor3000\relax}%
\providecommand \BibitemShut  [1]{\csname bibitem#1\endcsname}%
\let\auto@bib@innerbib\@empty
\bibitem [{\citenamefont {Voigt}\ \emph {et~al.}(2014)\citenamefont {Voigt}, \citenamefont {Ahrens}, \citenamefont {Heinrich}, \citenamefont {Thompson},\ and\ \citenamefont {Gruetzner}}]{voigt2014improved}%
  \BibitemOpen
  \bibfield  {author} {\bibinfo {author} {\bibfnamefont {A.}~\bibnamefont {Voigt}}, \bibinfo {author} {\bibfnamefont {G.}~\bibnamefont {Ahrens}}, \bibinfo {author} {\bibfnamefont {M.}~\bibnamefont {Heinrich}}, \bibinfo {author} {\bibfnamefont {A.}~\bibnamefont {Thompson}},\ and\ \bibinfo {author} {\bibfnamefont {G.}~\bibnamefont {Gruetzner}},\ }\bibfield  {title} {\bibinfo {title} {Improved adhesion of novolac and epoxy based resists by cationic organic materials on critical substrates for high volume patterning applications},\ }in\ \href@noop {} {\emph {\bibinfo {booktitle} {Advances in Patterning Materials and Processes XXXI}}},\ Vol.\ \bibinfo {volume} {9051}\ (\bibinfo {organization} {SPIE},\ \bibinfo {year} {2014})\ pp.\ \bibinfo {pages} {382--390}\BibitemShut {NoStop}%
\bibitem [{\citenamefont {Park}\ \emph {et~al.}(2018)\citenamefont {Park}, \citenamefont {Kim}, \citenamefont {Roh}, \citenamefont {Choi},\ and\ \citenamefont {Cha}}]{park2018simple}%
  \BibitemOpen
  \bibfield  {author} {\bibinfo {author} {\bibfnamefont {H.~W.}\ \bibnamefont {Park}}, \bibinfo {author} {\bibfnamefont {H.}~\bibnamefont {Kim}}, \bibinfo {author} {\bibfnamefont {J.~H.}\ \bibnamefont {Roh}}, \bibinfo {author} {\bibfnamefont {J.-K.}\ \bibnamefont {Choi}},\ and\ \bibinfo {author} {\bibfnamefont {K.-R.}\ \bibnamefont {Cha}},\ }\bibfield  {title} {\bibinfo {title} {Simple and cost-effective method for edge bead removal by using a taping method},\ }\href@noop {} {\bibfield  {journal} {\bibinfo  {journal} {Journal of the Korean Physical Society}\ }\textbf {\bibinfo {volume} {73}},\ \bibinfo {pages} {1473} (\bibinfo {year} {2018})}\BibitemShut {NoStop}%
\bibitem [{\citenamefont {Xie}(2025)}]{SiCNBcode}%
  \BibitemOpen
  \bibfield  {author} {\bibinfo {author} {\bibfnamefont {A.}~\bibnamefont {Xie}},\ }\href {https://www.flexcompute.com/tidy3d/kjFIWnQUtc} {\bibinfo {title} {{4H-SiC 1D Photonic Crystal Cavities}}} (\bibinfo {year} {2025})\BibitemShut {NoStop}%
\bibitem [{\citenamefont {Jin}(2025)}]{SiCTCcode}%
  \BibitemOpen
  \bibfield  {author} {\bibinfo {author} {\bibfnamefont {C.}~\bibnamefont {Jin}},\ }\href {https://www.flexcompute.com/tidy3d/fggro9Y3gb} {\bibinfo {title} {{4H-SiC 1D Photonic Crystal Cavities With Tapered Fiber Interface}}} (\bibinfo {year} {2025})\BibitemShut {NoStop}%
\bibitem [{\citenamefont {Dietz}\ \emph {et~al.}(2023)\citenamefont {Dietz}, \citenamefont {Jiang}, \citenamefont {Day}, \citenamefont {Bhave},\ and\ \citenamefont {Hu}}]{dietz2023spin}%
  \BibitemOpen
  \bibfield  {author} {\bibinfo {author} {\bibfnamefont {J.~R.}\ \bibnamefont {Dietz}}, \bibinfo {author} {\bibfnamefont {B.}~\bibnamefont {Jiang}}, \bibinfo {author} {\bibfnamefont {A.~M.}\ \bibnamefont {Day}}, \bibinfo {author} {\bibfnamefont {S.~A.}\ \bibnamefont {Bhave}},\ and\ \bibinfo {author} {\bibfnamefont {E.~L.}\ \bibnamefont {Hu}},\ }\bibfield  {title} {\bibinfo {title} {Spin-acoustic control of silicon vacancies in 4h silicon carbide},\ }\href@noop {} {\bibfield  {journal} {\bibinfo  {journal} {Nature Electronics}\ }\textbf {\bibinfo {volume} {6}},\ \bibinfo {pages} {739} (\bibinfo {year} {2023})}\BibitemShut {NoStop}%
\end{thebibliography}%

\end{document}


\title{Supplementary Information for "A Suspended 4H-Silicon Carbide Membrane Platform for Defect Integration into Quantum Devices"}

\date{\today}
\maketitle

\tableofcontents
\clearpage

\section{Additional Fabrication Methods}

\begin{figure}[ht!]
    \centering
    \includegraphics[width=1\linewidth]{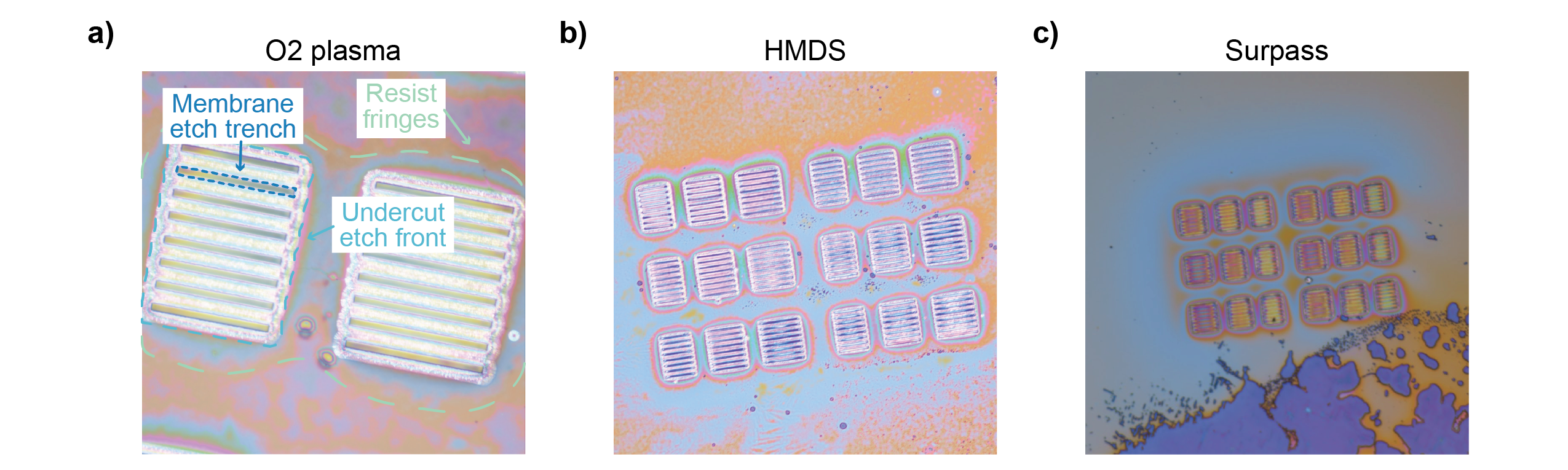}
    \caption{Optical images of ZEP520a spun onto suspended membrane samples with various surface pre-treatments. Images taken before baking. a) Representative sample treated with $O_2$ plasma. Various parameters were tested, ranging from: 100W-500W, 40 sccm, 15-300 sec. b) Representative sample treated with hexamethyldisilizane (HMDS). Two application methods were tested: vapor pressure oven and spinning (2k rpm, bake at 115$^\circ$C for 1 min). c) Sample treated with Surpass3000 \cite{voigt2014improved} (immersed for 60 sec, rinsed off in DI water), a cationic priming treatment which removes ionic contaminants and leaves the surface with a slight positive charge.}
    \label{SI1}
\end{figure}

Initially, ZEP520a application onto suspended membranes was inconsistent. Since anisole, the ZEP520a solvent, is only weakly polar, we tested both hydrophobic and hydrophilic surface pre-treatments (Figure S\ref{SI1}).  

\begin{figure}[h!]
    \centering
    \includegraphics[width=1\linewidth]{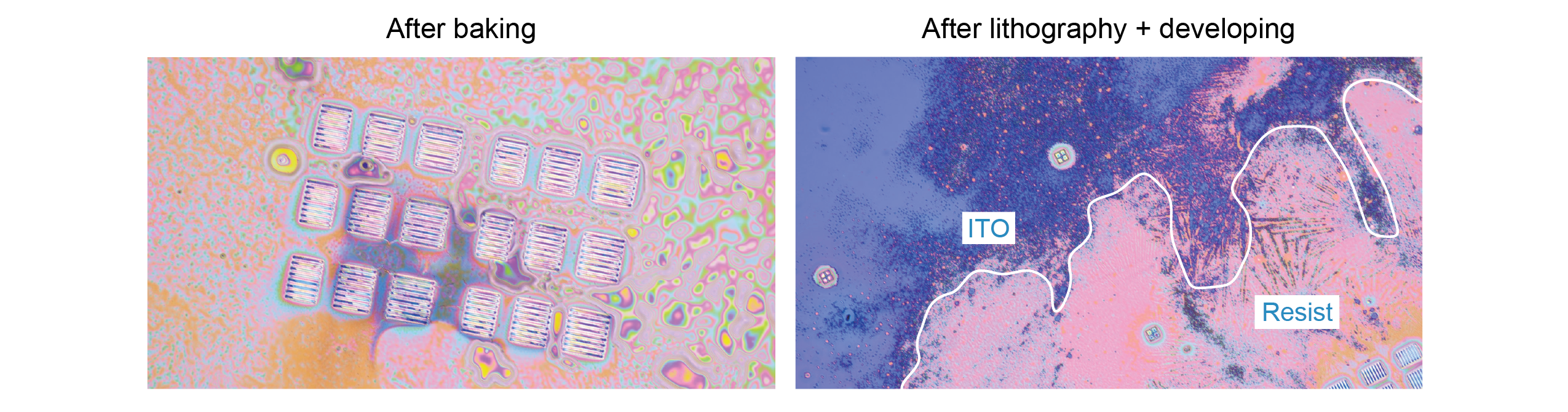}
    \caption{Optical images of same sample with suspended membranes after baking (left) and after lithography and development (right). Purple/blue film is exposed ITO resulting from resist delamination.}
    \label{SI2}
\end{figure}

The above mentioned surface treatments methods ultimately proved inconsistent in providing successful resist application. Though various combinations of the aforementioned methods were tested (e.g. $O_2$ plasma treatment followed by HMDS), none proved consistently helpful. The adhesion issues were further exacerbated during the high-temperature bake, where the resist buckled after initial attempts at baking (180$^\circ$C 3 min) and delaminated during development even on areas not exposed to the electron beam (Figure S\ref{SI2}). 

\newpage

\begin{figure}[h!]
    \centering
    \includegraphics[width=1\linewidth]{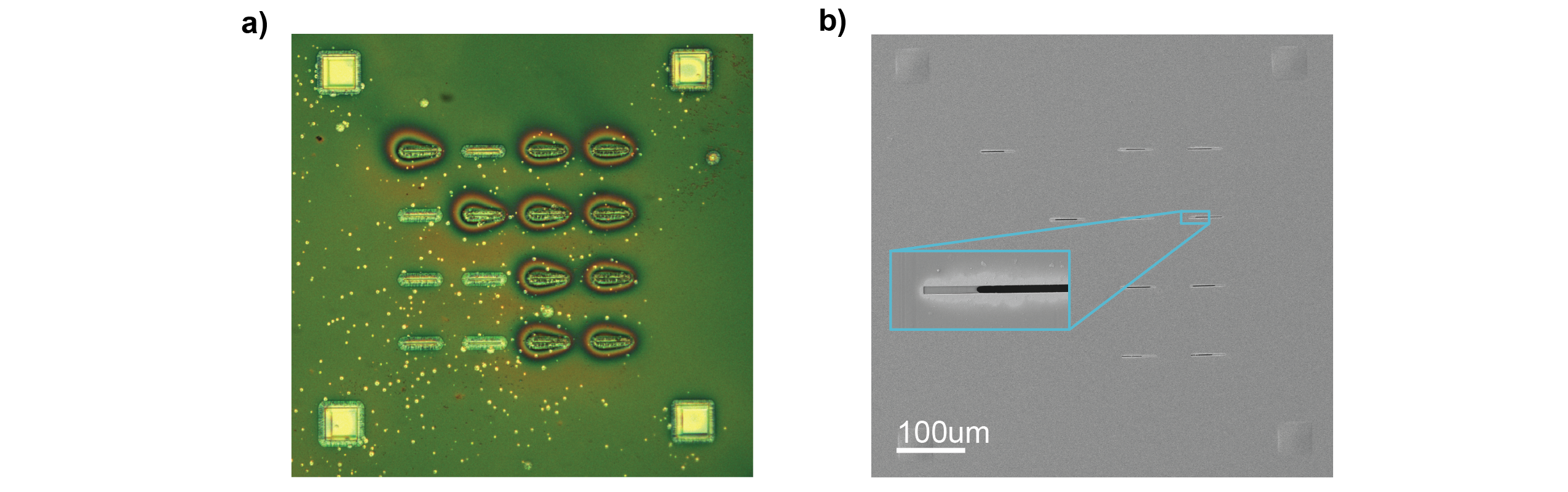}
    \caption{a) Optical image of ZEP520a spun on partially undercut membranes before baking. Fringes are indicative of edge beading, as seen in the SEM image. b) SEM image of the same sample in (a). Inset shows edge beading where resist does not evenly cover membrane surface. 10 nm of 80/20 Pt/Pd was sputtered on to prevent charging during imaging.}
    \label{SI3}
\end{figure}

Additionally, the small sample size (SiC chips were diced to 6x6 mm) and etch trench features increased edge beading, further complicating resist application (Figure S\ref{SI3}). 

\begin{figure}[h!]
    \centering
    \includegraphics[width=1\linewidth]{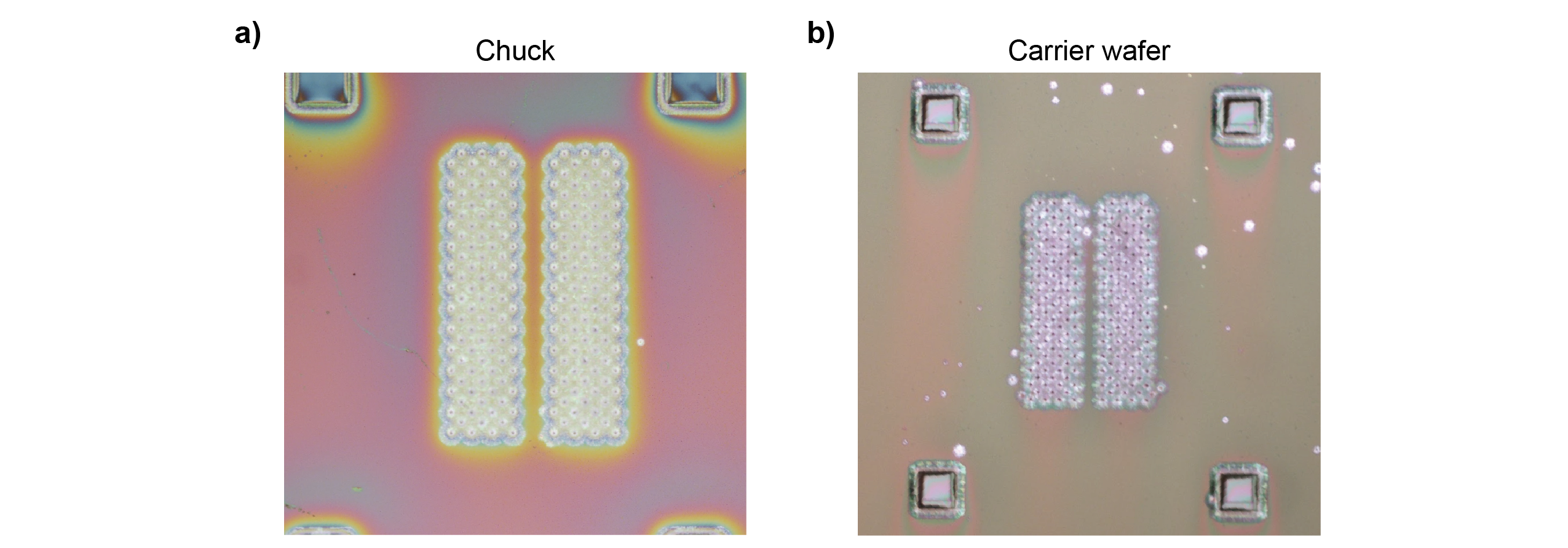}
    \caption{ZEP on a suspended sample spun on a) spinner chuck and b) on a carrier wafer. The same etch trench geometry was used for both samples.}
    \label{SI4}
\end{figure}

Ultimately, we found that a combination of lowering the baking temperature to 115$^\circ$C, and spinning the samples on a carrier wafer to prevent edge beading (Figure S\ref{SI4}) \cite{park2018simple}, provided the most reliable method for successful resist application. 

\newpage 

\section{Computational Design of Devices}

\subsection{1D Photonic Crystal Cavities}
Flexcompute Tidy3D was used to design for 1D PhCs with and without a waveguide interface. The associated code can be found in \cite{SiCNBcode} and \cite{SiCTCcode}.

\subsection{Phononic Cavity}
\begin{figure}[h!]
    \centering
    \includegraphics[width=1\linewidth]{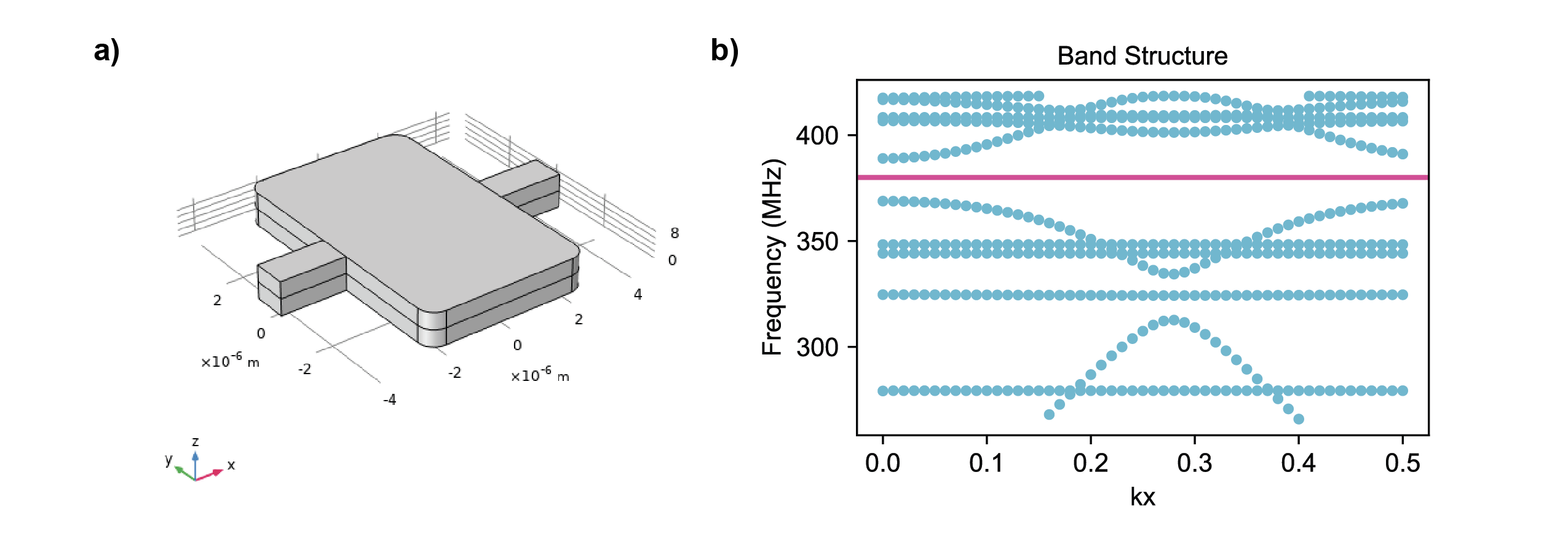}
    \caption{a) Mirror cell of block tether cavity in COMSOL. b) Band structure of mirror cell. Pink line is 380MHz.}
    \label{SI5}
\end{figure}

COMSOL Multiphysics was used to design the phononic resonator (Figure S\ref{SI5}). A bandgap surrounding 380 MHz was targeted to couple to acoustic transitions within the $V_{Si}^-$ ground state under low magnetic fields \cite{dietz2023spin}.

\section{Measurement Methods}

\subsection{Photoluminescence}
For 1D photonic crystal cavities, room temperature spectroscopy was performed with a LabRam Evolution Horiba Multiline Raman Spectrometer (50X LWD objective, 600 gr/mm grating, 2 sec acquisition time). Devices were excited using 532 nm (28 mW) to target surface state emission. 

\subsection{Cavity Reflection}
For tapered fiber cavities, room temperature cavity reflection measurements were performed with white light generated from NKT Photonics SuperK Extreme and sent in via a tapered fiber to the device. Spectra were taken with Acton SpectraPro-500i (1200 groves $mm^{-1}$, 800 nm blaze grating) and a Princeton Instruments visible 1340 x 400 pixels back-illuminated camera.

\bibliography{Ref.bib}